\documentclass[10pt]{article}
\usepackage[top=1in, bottom=1in, left=1in, right=1in]{geometry}
\usepackage{amsmath, amssymb, amsthm}
\usepackage{color, xcolor, colortbl}
\usepackage{graphics, subfig, overpic}
\usepackage{hyperref}
\usepackage{epstopdf}
\usepackage{srcltx}
\usepackage{algorithm, algpseudocode}
\usepackage[capitalize,nameinlink]{cleveref} 
\crefname{equation}{}{Equations}
\Crefname{equation}{Eq.}{Equations}
\usepackage{ulem}

\graphicspath{{images/}}
\numberwithin{equation}{section}

\newtheorem{remark}{Remark}

\newcommand\bbR{\mathbb{R}}
\newcommand\bbZ{\mathbb{Z}}
\newcommand\bbN{\mathbb{N}}

\newcommand\bbS{\mathbb{S}}

\newcommand\cE{\mathcal{E}}
\newcommand\cG{\mathcal{G}}
\newcommand\cL{\mathcal{L}}
\newcommand\cU{\mathcal{U}}
\newcommand\Id{\mathcal{I}}

\newcommand\dd{\,\mathrm{d}} 
\newcommand\pd[2]{\dfrac{\partial {#1}}{\partial {#2}}} 
   
\newcommand\sffont[1]{{\sf{#1}}}

\newcommand\Convone{{\sffont{Conv1d }}}
\newcommand\Convtwo{{\sffont{Conv2d }}}
\newcommand\CNNtwo{{\sffont{CNN2d}}}
\newcommand\CNNthree{{\sffont{CNN3d}}}
\newcommand\encoding{{\sffont{Encoding}}}
\newcommand\decoding{{\sffont{Decoding}}}
\newcommand\BCR{{\text{\sffont{BCR-Net}}}}
\newcommand\concatenate{{\sffont{Concatenate}}}
\newcommand\first{{\sffont{1d}}}
\newcommand\snd{{\sffont{2d}}}
\newcommand\trd{{\sffont{3d}}}

\newcommand\eps{\varepsilon}

\newcommand\diag{\mathrm{diag}\,}
\newcommand\NN{\mathrm{NN}}
\newcommand\T{\mathsf{T}}
\newcommand\cnn{\mathrm{cnn}}
\newcommand\Schrodinger{{Schr{\"o}dinger} }

\title{
  Solving Electrical Impedance Tomography with Deep Learning
}

\date{}
\author{
  Yuwei Fan  \thanks{Department of Mathematics, Stanford University, Stanford, CA 94305. 
    Email: {\tt ywfan@stanford.edu}},~~
  Lexing Ying  \thanks{Department of Mathematics and ICME, Stanford University, Stanford, CA 94305.
    Email: {\tt lexing@stanford.edu}}
}

\newcommand\delete[1]{}
\newcommand\add[1]{#1}
\newcommand\revised[2]{#2}

\renewcommand{\div}{\mathrm{div}}
\newcommand{\grad}{\nabla}
\newcommand{\p}{\partial}

\begin{document}
\maketitle

\begin{abstract}
  This paper introduces a new approach for solving electrical impedance tomography (EIT) problems
  using deep neural networks. The mathematical problem of EIT is to invert the electrical
  conductivity from the Dirichlet-to-Neumann (DtN) map. Both the forward map from the electrical
  conductivity to the DtN map and the inverse map are high-dimensional and nonlinear. Motivated by
  the linear perturbative analysis of the forward map and based on a numerically low-rank property,
  we propose compact neural network architectures for the forward and inverse maps for both 2D and
  3D problems.  Numerical results demonstrate the efficiency of the proposed neural networks.
\end{abstract}


\section{Introduction}\label{sec:intro}

Electrical impedance tomography (EIT) is the problem of determining the electrical conductivity
distribution of an unknown medium by making voltage and current measurements at the boundary of the
object. As a radiation-free imaging technique, EIT allows repeated, non-invasive measurements of
regional changes in the object; thus it has been used as a monitoring tool in a variety of
applications in critical care medicine, for instance, monitoring of ventilation distribution
\cite{victorino2004imbalances}, assessment of lung overdistension \cite{lindgren2007regional} and
detection of pneumothorax\cite{costa2008real}, and many industrial applications
\cite{wang2015industrial}.

\paragraph{Background.}
At the center of the mathematical formulations of EIT is the Dirichlet-to-Neumann (DtN) map, a
critical object in the analysis of elliptic partial differential equations that plays a significant
role in the classical Calder{\'o}n problem \cite{calderon2006inverse, uhlmann2009electrical,
  behrndt2015dirichlet}.

The governing equation of EIT, or equivalently the inverse conductivity problem, is
\begin{equation}\label{eq:divergence}
  \begin{aligned}
    -\div(\gamma(x)\grad \phi(x)) = 0,\quad &\text{ in } \Omega\subset\bbR^d,\\
    \phi(x) = \psi(x),\quad                 &\text{ on } \p \Omega,
  \end{aligned}
\end{equation}
where \add{$\Omega$ is a bounded Lipschitz domain,} $\phi(x)$ is the voltage, $\gamma(x)>0$ is the
conductivity distribution, and $\psi(x)$ is the voltage applied on the boundary. The corresponding
DtN map is defined by
\[
  \Lambda_\gamma: H^{\frac{1}{2}}(\p\Omega)\to H^{-\frac{1}{2}}(\p\Omega),\quad
  \psi(x)\mid_{\p\Omega} \to\gamma(x)\pd{\phi(x)}{n(x)}\mid_{\p\Omega},
\]
where $n(x)$ is the outer normal vector. \add{Here $H^{\frac{1}{2}}(\p\Omega)$ is the space of
$L^2(\p\Omega)$ functions that are traces of functions in $H^1(\Omega)$ and
$H^{-\frac{1}{2}}(\p\Omega)$ is its dual. We refer the readers to \cite{sylvester1990dirichlet} for
more details of the DtN map.}

A closely related inverse conductivity problem involves the DtN map of the \Schrodinger equation at
zero energy \cite{sylvester1990dirichlet}, which takes the following form
\begin{equation}\label{eq:schrodinger}
  \begin{aligned}
    (-\Delta+\eta(x))u(x) = 0, \quad &\text{ in } \Omega,\\
    u(x) = f(x),\quad                &\text{ on } \p \Omega.
  \end{aligned}
\end{equation}
The DtN map for the \Schrodinger equation is then defined by
\begin{equation}\label{eq:DtN}
  \Lambda_\eta: H^{\frac{1}{2}}(\p\Omega)\to H^{-\frac{1}{2}}(\p\Omega),\quad
  f(x)\mid_{\p\Omega} \to\pd{u(x)}{n(x)}\mid_{\p\Omega}.
\end{equation}

These two DtN maps $\Lambda_\eta$ and $\Lambda_\gamma$ are closely related. If $\phi$ is the
solution of \cref{eq:divergence}, then $u=\sqrt{\gamma}\phi$ is the solution of
\cref{eq:schrodinger} with $\eta = \frac{\Delta \sqrt{\gamma}}{\sqrt{\gamma}}$ and
$f=\sqrt{\gamma}\psi$. Moreover, $\Lambda_\eta =
\gamma^{-1/2}\Lambda_\gamma\gamma^{-1/2}+\dfrac{1}{2\gamma}\pd{\gamma}{n}$. Actually, the two maps
$\Lambda_\eta$ and $\Lambda_\gamma$ carry the same information and they can be determined from each
other \cite{sylvester1990dirichlet}. This paper shall focus on the DtN map $\Lambda_\eta$ for the
\Schrodinger equation.  All the results can be extended to the DtN map $\Lambda_\gamma$ without many
difficulties.

Since the DtN map $\Lambda_\eta$ is linear \cite{behrndt2015dirichlet} for a fixed $\eta$, there
exists a distribution kernel $\lambda_\eta(r,s)$ for $r,s\in\p\Omega$ such that
\begin{equation}
  (\Lambda_\eta f)(r)=\pd{u}{n}(r)=\int_{\p\Omega}\lambda_\eta(r,s)f(s)\dd S(s).
\end{equation}
The forward problem for the DtN map is that, given $\eta(x)$, to solve for the kernel
$\lambda_\eta(r,s)$, i.e., $\eta\to \lambda_\eta$.

The task of the inverse problem is to recover $\eta(x)$ in $\Omega$ based on the observation data,
which is typically a collection of pairs $(f,\Lambda_\eta f)$ of the Dirichlet boundary condition
$f$ and the corresponding Neumann data $\Lambda_\eta f$. Under the assumption that the Dirichlet
boundary condition is sufficiently sampled, it is possible to assume that the kernel $\lambda_\eta$
is known and, therefore, the inverse problem is to recover $\eta$ from $\lambda_\eta$, i.e.,
$\lambda_\eta\to \eta$. Since $\lambda_\eta(r,s)\mid_{r,s\in\p\Omega}$ is a function of $2(d-1)$
variables while $\eta(x)$ is a function of $d$ variables, the inverse problem is not solvable if
$d=1$ due to a simple dimension counting. For $d\geq2$, in principle, the solution of the inverse
problem exists and is unique under certain conditions \cite{uhlmann2016dirchlet}. However, due to
the elliptic nature of EIT, the inverse problem is severely ill-conditioned
\cite{alessandrini1988stable,alessandrini1997examples, allers1991stability, borcea2002electrical}
even for $d\ge 2$.

Numerical solution of the forward and inverse problems can be challenging. The forward problem is a
map from a $d$ dimensional function to a $2(d-1)$ dimensional function. For 3D problems, computing
and representing the whole DtN map $\Lambda_\eta$ for a fixed $\eta$ can be quite expensive. For the
inverse problem, the inverse map $\Lambda_\eta \to \eta$ is numerically unstable
\cite{alessandrini1988stable,alessandrini1997examples, allers1991stability, borcea2002electrical}
due to ill-conditioning. In order to avoid instability, an application-dependent regularization term
is often required in order to stabilize the inverse problem, see, for instance,
\cite{hanke1997regularizing, chan1999nonlinear, kaipio1999inverse}. Algorithmically, the inverse
problem is usually solved with iterative methods
\cite{hanke1997regularizing,haber2000optimization,borcea2001nonlinear, borcea2002electrical}, which
often requires a significant number of iterations.

In the last few years, deep neural networks (DNNs) have achieved great successes in computer vision,
image processing, speech recognition, and many other artificial intelligence applications
\cite{Hinton2012,Krizhevsky2012,goodfellow2016deep,MaSheridan2015,Leung2014,SutskeverNIPS2014,leCunn2015,SCHMIDHUBER2015}.
More recently, methods based on DNNs have also been applied to solving PDEs
\cite{khoo2017solving,berg2017unified,han2018solving,fan2018mnn,fan2018mnnh2,Araya-Polo2018,Raissi2018,fan2019bcr,khoo2018switchnet}.
These attempts can be classified into two categories. The first category
\cite{rudd2015constrained,carleo2017solving,han2018solving,khoo2019committor,weinan2018deep} aims to
represent the solutions of high-dimensional PDEs with DNNs (rather than the classical methods such
as finite element and finite difference methods). The second category
\cite{long2018pde,han2017deep,khoo2017solving,fan2018mnn,fan2018mnnh2,fan2019bcr,khoo2018switchnet,li2019variational,bar2019unsupervised}
works with parameterized PDE problems and uses the DNNs to represent the map from the
high-dimensional parameters of the PDE to the solution of the PDE.

\paragraph{Contributions.}

Deep neural networks have several advantages when applied to solve the forward and inverse problems.
For the forward problem, since applying neural network to input data can be carried out rapidly due
to novel software and hardware architectures, the forward problem can be significantly accelerated
when the forward map is represented with a DNN. For the inverse problem, the choices of the solution
algorithm and the regularization term are two critical issues. Fortunately, deep neural networks can
help in both aspects. First, concerning the solution algorithm, due to its flexibility in
representing high-dimensional functions, DNN can potentially be used to approximate the full inverse
map, thus avoiding the iterative solution process. Second, concerning the regularization term,
recent work in machine learning shows that DNNs often can automatically extract features from the
data and offer a data-driven regularization prior.

This paper applies the deep learning approach to the EIT problem by representing the inverse map
from $\Lambda_\eta$ to $\eta$ using a novel neural network architecture. The motivation of the new
architecture comes from a perturbative analysis of the linear approximation of both the forward and
inverse maps of the EIT problem. The analysis shows that the maps between $\eta$ and $\Lambda_\eta$
are locally numerically low-rank after a reparameterization of the DtN map $\Lambda_\eta$. This
observation allows us to reduce the map between the $d$-dimensional $\eta$ and $2(d-1)$-dimensional
$\Lambda_\eta$ to a map between two (quasi) $(d-1)$-dimensional functions. Being
translation-invariant and global, this new map is represented with the recently proposed BCR-Net
\cite{bcr}, which is a multiscale neural network based on the nonstandard form of the wavelet
decomposition. This neural network architecture is used to approximate both the forward and inverse
maps. For the test problems being considered, the resulting neural networks have only $10^4\sim10^5$
parameters for the 2D case and $10^5\sim10^6$ parameters for the 3D case, thanks to the dimension
reduction and the compact structure of the BCR-Net. The rather small number of parameters allow for
training on rather limited data sets, which are often the case for EIT problems.

\paragraph{Organization.}
This rest of the paper is outlined as follows. The mathematical background on the DtN map is studied
in \cref{sec:math}. The design and architecture of the DNNs of the forward and inverse maps for the
2D case are discussed in \cref{sec:2d}, along with numerical tests. The result is extended to the 3D
case in \cref{sec:3d}.

\section{Mathematical analysis of the DtN map}\label{sec:math}

This section summarizes the necessary mathematical background of the DtN map. Let us denote
$\cL=-\Delta+\eta$ and $\cG=\cL^{-1}$ with $\cG f(x)=\int_{\Omega}G(x,y)f(y)\dd y$, where $G$ is the
Green function of the operator $\cL$ with the Dirichlet boundary condition. An application of the
divergence theorem shows that 
\begin{equation}
    0 = \int_{\p \Omega}\pd{u}{n(y)}(y)G(x,y) \dd S(y)
    = \int_{\Omega} \div_y(\grad_y u(y) \cdot G(x,y))\dd y 
    = \int_{\Omega} \left(\Delta_y u \cdot G + \grad_y G \grad_y u\right)\dd y.
\end{equation}
Analogously, a second application of the divergence theorem to the above result leads to
\begin{equation}\label{eq:schr_G}
    \begin{aligned}
        \int_{\p \Omega}\pd{G}{n(y)}(x,y) f(y) \dd S(y) 
        & = \int_{\Omega} \div_y(\grad_y G(x,y) \cdot u(y))\dd y \\
        & = \int_{\Omega} \left(\Delta_y G(x,y) \cdot u(y) + \grad_y G(x,y)\grad_y u(y) \right) \dd y\\
        & = \int_{\Omega}\left( \Delta_y G(x,y)\cdot u(y) -\Delta_yu(y)\cdot G(x,y) \right)\dd y\\
        & = \int_{\Omega}\left(-\left( -\Delta_y+\eta(y)\right) G(x,y)\cdot u(y) 
        + \left(-\Delta_y+\eta(y)\right)u(y) \cdot G(x,y) \right)\dd y\\
        &= -u(x).
    \end{aligned}
\end{equation}
Here the last equality uses the fact that $G$ is the Green function of $\cL=-\Delta+\eta$ and $u$ is
the solution of \cref{eq:schrodinger}. Taking the normal derivative of two sides of
\cref{eq:schr_G} with respect to $n(x)$ for $x\in\p\Omega$ gives rise to
\begin{equation}\label{eq:DtN_schr}
  \pd{u}{n}(x) = -\int_{\p\Omega}\pd{^2 G}{n(x)n(y)}(x,y)f(y)\dd S(y),\quad x\in\p\Omega,
\end{equation}
which describes the kernel of the DtN map $\Lambda_{\eta}(\psi)$ in terms of the Green function $G$:
\begin{equation}\label{eq:lambda}
  \lambda_\eta(r, s) = -\pd{^2 G}{n(r)n(s)}(r,s),\quad r,s\in\p\Omega.
\end{equation}
In order to avoid confusion, we use $r,s$ to represent the points on the boundary and $p,q$ for the
points in the domain hereafter.

In order to understand how the DtN map depends on the potential $\eta$, we conduct a perturbative
analysis of the map from $\eta$ to $\lambda_\eta$ for $\eta>0$ close to a fixed $\eta_0$. For
simplicity, assume $\eta_0=0$. Let us introduce \revised{$\cE=-\diag(\eta)$}{$\cE=-\eta\Id$ with
  $\Id$ the identity operator,} $\cL_0=-\Delta$, and $\cG_0=\cL_0^{-1}$ (with kernel denoted by
$G_0$) as the Green function of $\cL_0$ with the Dirichlet boundary condition. When $\eta>0$ is
sufficiently small, $\cG$ can be expanded via a Neumann series
\begin{equation}\label{eq:expansion_G}
  \cG = (\cL_0-\cE)^{-1}=\cG_0 + \cG_0\cE \cG_0 + \cG_0\cE\cG_0\cE\cG_0+\dots.
\end{equation}
By introducing $\lambda_0(r,s) = \lambda_\eta(r,s)\mid_{\eta=\eta_0}$, which can be calculated by the
knowledge of the background case $\eta=\eta_0$, it is equivalent to focus on the difference
$\lambda_\eta-\lambda_0$ (often called {\em difference imaging}, see \cite{eit2003review} for
details), which is also the kernel of  $\cG-\cG_0$. 
For a sufficiently small $\eta$, the operator $\cG-\cG_0$ can be approximated by its first term
$\cG_0\cE \cG_0$, which is linear in $\cE$. Using the fact that
\revised{$\cE=-\diag(\eta)$}{$\cE=-\eta\Id$} leads to the following approximation for the {\em
difference DtN map} $\mu$,
\begin{equation}\label{eq:V2lambda}
  \mu(r,s):=(\lambda_\eta-\lambda_0)(r,s) = -\pd{^2(G-G_0)}{n(r)\p n(s)}(r,s) 
  \approx \int_{\Omega} \left( \pd{G_0}{n(r)}(r, p)\pd{G_0}{n(s)}(p,s) \right) \eta(p)\dd p,
\end{equation}
which serves as the motivation of the design of the NN architectures.

\section{Neural network for the 2D case}\label{sec:2d}

\begin{figure}[htb]
  \centering
  \subfloat[One-sided detection]{
    \includegraphics[width=0.49\textwidth,page=3]{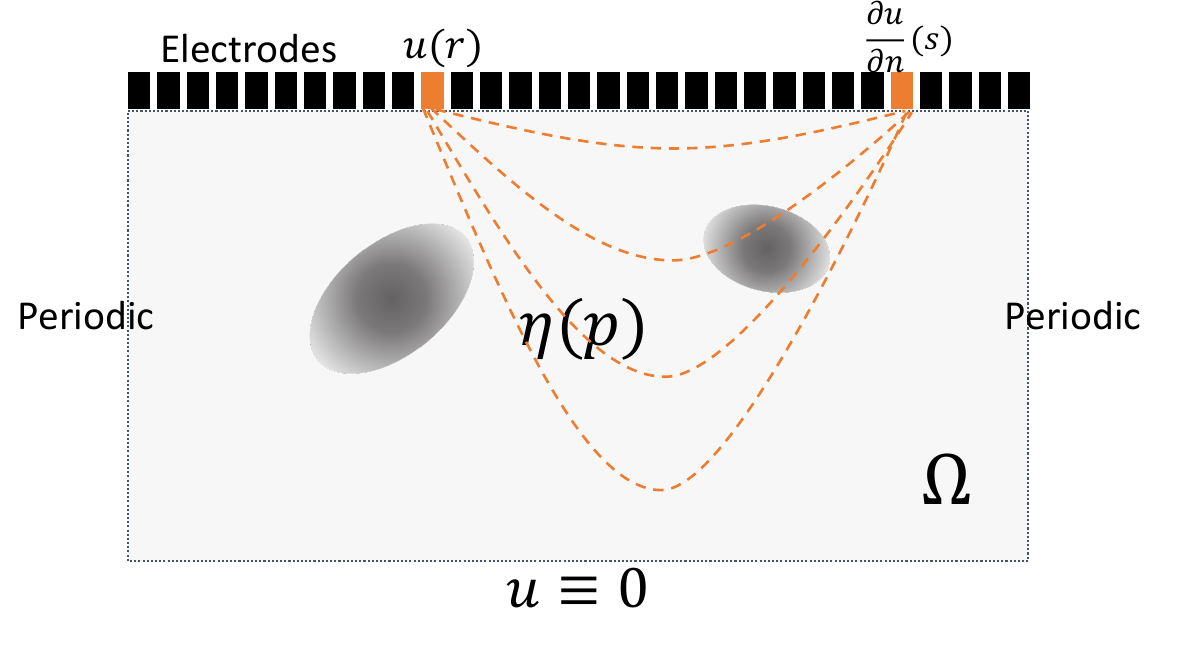}
  }
  \subfloat[Two-sided detection]{
    \includegraphics[width=0.49\textwidth,page=4]{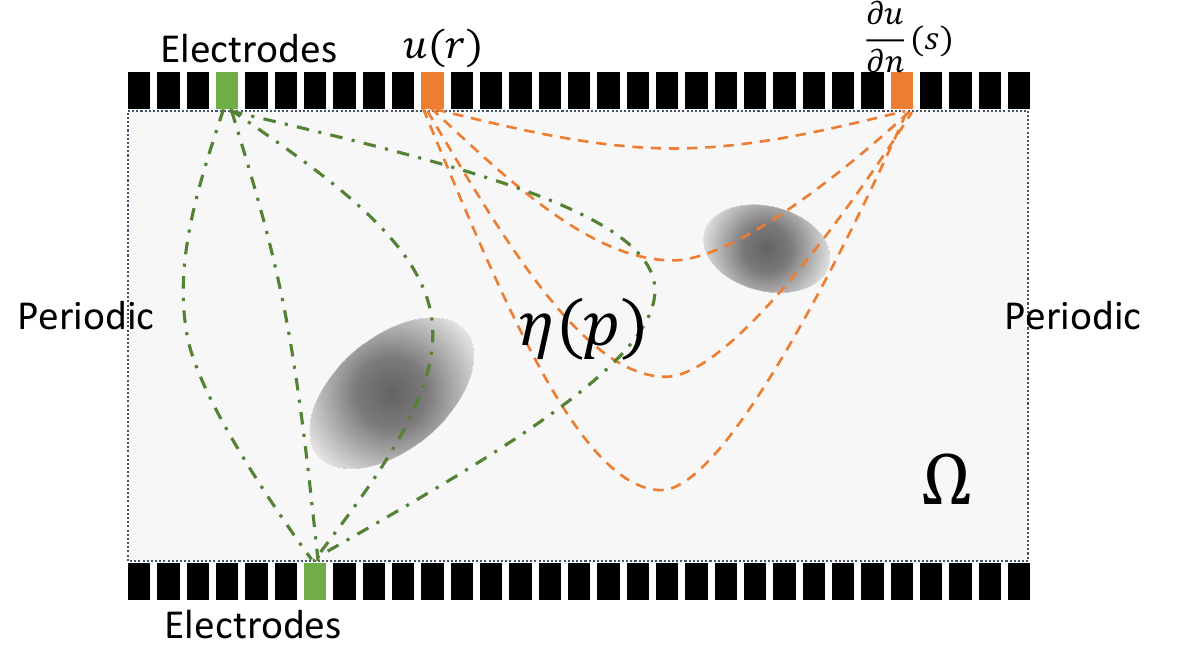}
  }
  \caption{\label{fig:domain}Illustration of the problem setup. In the one-sided detection, the
    electrodes are only placed on the top and the zero Dirichlet boundary condition is applied on
    the bottom. In the two-sided detection, the electrodes are placed on both the top and bottom.
    In both cases, periodic boundary conditions are applied on the left and right boundaries.}
\end{figure}

Consider the domain $\Omega=[0,1]\times [-Z,Z]$ where $Z$ is a fixed constant. The periodic boundary
condition is specified at the left and right boundaries for simplicity. As illustrated in
\cref{fig:domain}, the electrodes are allowed to be placed on either only the top boundary
(one-sided detection) or both the top and bottom boundaries (two-sided detection). For the one-sided
detection, the zero Dirichlet boundary condition is assumed at the bottom for simplicity, though
other boundary conditions are also relevant. In what follows, we shall first consider the forward
and inverse maps for the one-sided detection. The architecture is then extended to the two-sided
detection case.

In most of the EIT problems, the electrical conductivity is known near the domain boundary. This
implies that there exists a constant $\delta>0$ such that $\eta(p)$ is supported in $[0,1]\times
[-(Z-\delta),Z-\delta]$.

\subsection{Forward map for the one-sided detection}\label{sec:forward}

For the one-sided detection, the DtN map is limited on the top boundary.  Let $r=(r_1, Z)$,
$s=(s_1,Z)$ and $p=(x,z)$, where $x$ is for the horizontal coordinate and $z$ is the depth
coordinate. The map \cref{eq:V2lambda} can be rewritten as
\begin{equation}\label{eq:approx}
    \mu( (r_1, Z), (s_1,Z)) \approx
    \int_{\Omega}\pd{G_0}{n(r)}( (r_1, Z), (x,z))
    \pd{G_0}{n(s)}( (s_1,Z),(x,z))\eta(x,z)\dd x\dd z.
\end{equation}
Note that $p$ and $q$ are used for the points in the domain $\Omega$, $r$ and $s$ for the points on
the boundary, and $x$ and $z$ for the horizontal and depth coordinates.

A key step for both analysis and architecture design is to introduce new horizontal variables $m$
and $h$ such that $r_1=m+h$ and $s_1=m-h$. Reparameterizing the difference DtN map $\mu$ with the
new variable yields
\begin{equation}\label{eq:def_kappa}
  \mu(m,h) :=\mu( (m+h,Z), (m-h,Z) ) \approx  \int_{\Omega}K(m,h,x,z)\eta(x,z)\dd x\dd z,
\end{equation}
with the kernel $K$ given by
\begin{equation}\label{eq:def_K}
    K(m,h,x,z) := \pd{G_0}{n}( (m+h, Z), (x,z)) \pd{G_0}{n}( (m-h,Z),(x,z)).
\end{equation}
Here $n=(0,1)$ and $\pd{G_0}{n}(\cdot,\cdot)$ is the directional derivative of $G_0$ in the first
variable. Noticing that $G_0$ is the Green function of the operator $-\Delta$ on the domain $\Omega$
with the periodic boundary condition on left and right and the Dirichlet boundary condition on top
and bottom, one can write down $G_0$ explicitly as \cite{pde2nd2001}
\begin{equation}\label{eq:G0}
  G_0(p,q) = \sum_{\ell\in\bbZ^2}\left(\Gamma(p-q+(\ell_1,2\ell_2Z)) - \Gamma(p-q^\ast+(\ell_1, 2\ell_2Z))\right),
\end{equation}
where $q^\ast=(q_1,2Z+q_2)$ and $\Gamma$ is the Green function of the operator $-\Delta$ on the
whole space $\bbR^2$. Since $G_0$ as the Green function for the case $\eta=\eta_0$ is
translation-invariant \add{in the horizontal direction},
\begin{equation}
  \pd{G_0}{n}( (m\pm h,Z),(x,z) ) = \pd{G_0}{n}( (\pm h, Z),(x-m,z)).
\end{equation}
For the rest of the discussion, it is convenient to treat $h$ and $z$ as parameters and introduce
\begin{equation*}
  \begin{aligned}
    &\pd{G_{0,\pm h, z}}{n}(x-m) := \pd{G_0}{n}( (\pm h, Z),(x-m,z), \quad &
    &\eta_z(x) :=\eta(x,z), \\
    &k_{h,z}(m) := \pd{G_{0,+h,z}}{n}(m)\pd{G_{0,-h,z}}{n}(m), &
    &\mu_h(m) := \mu(m,h).
  \end{aligned}
\end{equation*}
With the new notations, \cref{eq:def_kappa} can be reformulated as
\begin{equation}\label{eq:kappa}
  \mu_h(m) \approx \int_{-Z}^{Z}(k_{h,z}*\eta_z)(m)\dd z
  = \int_{-(Z-\delta)}^{Z-\delta}(k_{h,z}*\eta_z)(m)\dd z,
\end{equation}
where the convolution is in $m$. The last equality holds due to the consideration that $\eta$ is
supported between $-(Z-\delta)$ and $Z-\delta$ in the depth direction.

\paragraph{Low-rank approximation and dimension reduction.}

A key observation is that
\begin{center}
  the kernel $k_{h,z}(m)$ is smooth in $h$ for $h\in[0,1]$ and $z \in
  (-(Z-\delta),Z-\delta)$.
\end{center}
An inspection of the definition of $K$ in \cref{eq:def_K} shows that $k_{h,z}(m)$ is only singular
when $z=Z$. Therefore, the kernel $k_{h,z}(m)$ is uniformly smooth for $h\in[0,1]$, $m\in[0,1]$, and
$z\in(-(Z-\delta),Z-\delta)$. The smoothness in the $h$ and $z$ variable indicates that $k_{h,z}(m)$
can be well-approximated in $h$ and $z$ by an approximation scheme with a small number of terms. To
simplify the discussion, assume without loss of generality that a stable interpolation scheme (such
as Chebyshev interpolation) is adopted.  By denoting the sets of interpolation points in the $h$
variable and $z$ variable as $\{\hat{h}\}$ and $\{\hat{z}\}$, such an interpolation reads
\begin{equation}\label{eq:lowrank}
  k_{h,z}(m) \approx \sum_{\hat{h}} \sum_{\hat{z}} R_{h,\hat{h}} k_{\hat{h},\hat{z}}(m) R_{z,\hat{z}},
\end{equation}
where $R_{h,\hat{h}}$ and $R_{z,\hat{z}}$ are the interpolation operators in the $h$ and $z$
variables, respectively.

This approximation for $k_{h,z}$ naturally implies an approximation for \cref{eq:kappa}
\begin{equation}\label{eq:lowrank_app}
  \mu_h(m) \approx 
  \sum_{\hat{h}} R_{h,\hat{h}} \left( \sum_{\hat{z}} k_{\hat{h},\hat{z}}*
  \left(\int_{-(Z-\delta)}^{Z-\delta}  R_{z,\hat{z}}\eta_z\dd z\right)\right)(m).
\end{equation}
Algorithmically, this approximation allows one to factorize the forward map into three steps:
\begin{enumerate}
\item Compress the two-dimensional function $\eta_z=\eta(x,z)$ to a set of one-dimensional function
  \[
  \tilde{\eta}_{\hat{z}}(x) := \int_{-(Z-\delta)}^{Z-\delta}R_{z,\hat{z}} \eta_z(x)\dd z;
  \]  
\item Convolve with $k_{\hat{h},\hat{z}}$ in the one-dimensional space to obtain
  \[
  \tilde{\mu}_{\hat{h}}(m) := \left( \sum_{\hat{z}} k_{\hat{h},\hat{z}}*\tilde{\eta}_{\hat{z}} \right)(m);
  \]
\item Interpolate the set of one-dimensional functions $\tilde{\mu}_{\hat{h}}(m)$ to a
  two-dimensional function
  \[
  \mu_h(m)=\sum_{\hat{h}} R_{h,\hat{h}} \tilde{\mu}_{\hat{h}}(m) .
  \]
\end{enumerate}
This effectively reduces the forward map to a number of 1D convolutions. This {\em dimension
reduction} in \cref{eq:lowrank_app} is fundamental in the construction of the neural network.

\add{
\begin{remark}
  The assumption that $\eta(p)$ is supported in $[0,1]\times [-(Z-\delta), Z-\delta]$ can be
  removed. Actually, we can split $[-Z,Z]$ into three intervals $[-Z, -(Z-\delta)]$, $[-(Z-\delta),
    Z-\delta]$ and $[Z-\delta, Z]$ with $\delta\ll Z$, and then study the property of the kernel
  $k_{h,z}(m)$ restricted to each interval one by one. Since $\delta \ll Z$, the low-rank
  approximation \cref{eq:lowrank_app} is still valid.
\end{remark}
}

\paragraph{Discretization.}
The analysis till now is in the continuous setting. A simple numerical treatment discretizes the
domain $\Omega$ by a uniform Cartesian grid, with the Laplacian approximated by a $5$-point central
difference scheme and the directional derivative on the boundary replaced by the one-sided
first-order difference. The numerical Green function is defined to be the inverse of the discrete
Laplacian operator with zero boundary condition. Let $N_r$ be the number of electrodes. \add{The DtN
  map is evaluated by solving \cref{eq:schrodinger} $N_r$ times with $f(x)$ set as a delta function
  at one electrode each time.} With a slight abuse of notation, the same letters are used to denote
the continuous kernels and their discretizations. The discrete version of \cref{eq:lowrank_app}
reads
\begin{equation}\label{eq:lowrank_discret}
  \mu_h(m) \approx
  \sum_{\hat{h}} R_{h,\hat{h}} \left( \sum_{\hat{z}} k_{\hat{h},\hat{z}}*
  \left(\sum_{z} R_{z,\hat{z}} \eta_z\right) \right)(m).
\end{equation}

\paragraph{Neural network architecture.}

The perturbative analysis shows that, if $\eta>0$ is sufficiently small, the forward map
$\eta\to\mu$ can be approximated by \cref{eq:lowrank_discret}. As detailed below, the three
steps of computing \cref{eq:lowrank_discret} can be naturally formulated as a neural network with
three modules:
\begin{itemize}
\item an {\em encoding} module that compresses the two-dimensional data $\eta$ to a set of
  one-dimensional data $\tilde{\eta}_{\hat{z}}$; 
\item an intermediate module that convolves $k_{\hat{h},\hat{z}}$ with the one-dimensional data
  $\tilde{\eta}_{\hat{z}}$ to obtain $\tilde{\mu}_{\hat{h}}$;
\item a {\em decoding} module that extends the set of one-dimensional data
  $\tilde{\mu}_{\hat{h}}$ to two-dimensional data $\mu$.
\end{itemize}

When $\eta$ fails to be sufficiently small, the linear approximation for the forward map
$\eta\to\mu$ is not accurate. In order to extend the neural network of \cref{eq:lowrank_discret} to
the nonlinear case, a straightforward solution is to include nonlinear activation functions and
increase the number of layers, for instance in \cite{fan2018mnn,fan2019bcr}. For simplicity, we
assume that the size $N_{\hat{z}}$ of $\{\hat{z}\}$ and the size $N_{\hat{h}}$ of $\{\hat{h}\}$ are
both equal to a constant parameter $c$.

\begin{algorithm}[htb]
  \begin{small}
    \begin{center}
      \begin{algorithmic}[1]
        \Require $c=N_{\hat{z}}=N_{\hat{h}}, n_{\cnn}\in\bbN$, $\eta\in\bbR^{N_{x}\times N_{z}}$
        \Ensure $\mu\in\bbR^{N_m\times N_{{h}}}$
        \State $\tilde{\eta} \leftarrow \encoding[c](\eta)$
        \State $\tilde{\mu} \leftarrow \BCR\first[c, n_{\cnn}](\tilde{\eta})$
        \State $\mu \leftarrow \decoding[N_h](\tilde{\mu})$
        \State \sffont{return} $\mu$
      \end{algorithmic}
    \end{center}
  \end{small}
  \caption{\label{alg:eitforward} Neural network architecture for the one-sided detection forward map
    $\eta\to \mu$.}
\end{algorithm}
\begin{figure}[htb]
  \centering
  \includegraphics[width=0.6\textwidth]{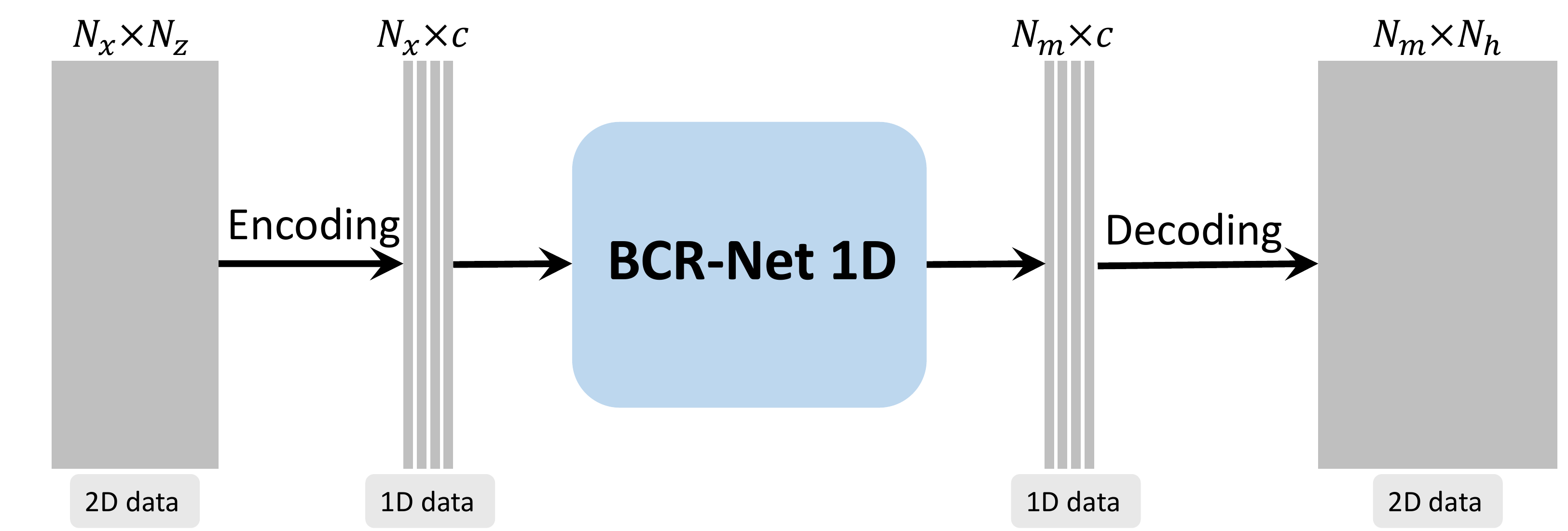}
  \caption{\label{fig:eitforward}Neural network architecture for the forward problem of the 
    one-sided detection.}
\end{figure}

The resulting neural network architecture for the forward map for the one-sided detection is
summarized in \cref{alg:eitforward} and illustrated in \cref{fig:eitforward}. Let us explain these
three components of the neural network one by one.
\begin{itemize}
\item Encoding module.  $\tilde{\eta} = \encoding[c](\eta)$ compresses the data
  $\eta\in\bbR^{N_{x}\times N_{z}}$ to $\tilde{\eta}\in\bbR^{N_x\times c}$ by compressing only in
  the $z$-dimension.
  It can be 
  implemented with a one-dimensional convolutional layer \Convone with window size $1$ and channel
  number $c$ by taking the second dimension of $\eta$ as channels. The linear activation function is
  sufficient for the \Convone layer used here.
\item Intermediate module. Since the kernel $k_{\hat{h},\hat{z}}$ for the linear case in
  \cref{eq:lowrank_discret} is a convolution, it can be implemented by a one-dimensional
  convolutional layer \Convone with window size $N_x$, channel number $c$ and linear
  activation function. For nonlinear case, a natural extension is to use multiple convolution
  layers and to add a nonlinear activation function such as a rectified-linear unit (ReLU) function
  after each layer.
  
  For problems with fine discretizations, a \delete{fully-connected} convolution layer \add{with
    window size $N_x$} might have many parameters. Recently, several multiscale NNs with fewer
  parameters have been proposed as an efficient alternative to
  \revised{fully-connected}{full-width} convolution layers. Examples include the ones based on
  hierarchical matrices in \cite{fan2018mnn,fan2018mnnh2} and the BCR-Net \cite{fan2019bcr}. Here,
  the BCR-Net is used to represent the intermediate module. BCR-Net is motivated by the data-sparse
  nonstandard wavelet representation of the pseudo-differential operators \cite{bcr}. It processes
  the information at different scale separately and each scale can be understood as a {\em local}
  convolutional neural network. The \revised{sub-network}{one-dimensional}
  $\tilde{\mu}=\BCR\first[c, n_{\cnn}](\tilde{\eta})$ maps $\tilde{\eta}\in\bbR^{N_x\times c}$ to
  $\tilde{\mu}\in\bbR^{N_x\times c}$, where the number of channels and layers in the local
  convolutional neural network in each scale are $c$ and $n_{\cnn}$, respectively. The readers are
  referred to \cite{fan2019bcr} for more details on the BCR-Net.
\item Decoding module.  $\mu=\decoding[N_h](\tilde{\mu})$ decodes the set of one-dimensional data
  $\tilde{\mu}\in\bbR^{N_m\times c}$ to the two-dimensional data $\mu\in\bbR^{N_m\times N_h}$. 
  \delete{It is the adjoint of the encoding module.}
  In the implementation, this decoding module is implemented by the one-dimensional convolutional
  layer \Convone with window size $1$, channel number $N_h$, and linear activation function.
\end{itemize}

\subsection{Inverse map for the one-sided detection} \label{subsec:inv1s}
The perturbative analysis shows that if $\eta$ is sufficiently small, the forward map can be
well-approximated by
\begin{equation}
  \mu \approx K \eta,
\end{equation}
which is the operator notation of the discretization \cref{eq:def_kappa}. Here, $\eta$ is a vector
indexed by $(x,z)$, $\mu$ is indexed by $(m,h)$, and $K$ is a matrix with rows indexed by $(m,h)$
and columns indexed by $(x,z)$. The usual filtered back-projection algorithm
\cite{holder2004electrical} takes the form
\[
\eta \approx (K^\T K + \eps I)^{-1} K^\T \mu,
\]
where $\eps$ is a regularization parameter.

Following the above discussion, the dimension reduction approximation applied to $K$ is also valid
for $K^\T$
\[
(K^\T \mu)_{z}(x) \approx 
\sum_{\hat{z}} R_{z,\hat{z}} \left( \sum_{\hat{z}} k_{\hat{h},\hat{z}}*
\left(\sum_{h} R_{h,\hat{h}} \mu_h\right) \right)(x).
\]
As a result, one obtains a similar three-step algorithm for applying $K^\T$ to $\mu$ and this
algorithm can also be formulated as a neural network with three modules:
\begin{itemize}
\item Encode from $\mu$ to $\tilde{\mu}_{\hat{h}} = \sum_{h} R_{h,\hat{h}} \mu_h$.
\item Convolve to form $\tilde{\eta}_{\hat{z}} = \sum_{\hat{h}} k_{\hat{h},\hat{z}} *\tilde{\mu}_{\hat{h}}$.
\item Decode from $\tilde{\eta}_{\hat{z}}$ to $(K^\T \mu)_{z} =
  \sum_{\hat{z}}R_{z,\hat{z}}\tilde{\eta}_{\hat{z}}$
\end{itemize}
\add{The part $(K^\T K+\eps I)^{-1}$ can be viewed as a post-processing of $K^\T \mu$.}  The
definition of $K$ \cref{eq:def_K} implies that the operator $(K^\T K + \eps I)$ is a convolution
operator. As a deconvolution operator, $(K^\T K + \eps I)^{-1}$ can also be implemented with a
convolution neural network.

\begin{figure}[htb]
  \centering
    \includegraphics[width=0.8\textwidth]{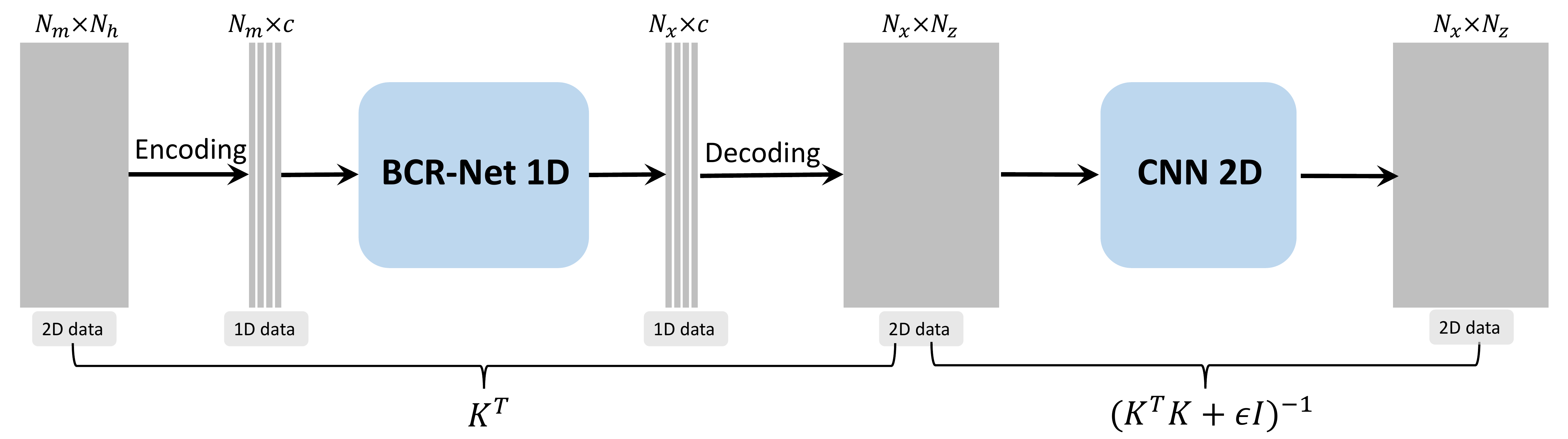}
  \caption{\label{fig:eitinverse}Neural network architecture for the inverse problem of the
    one-sided detection.}
\end{figure}

Combining these two components suggests that for the inverse map a suitable architecture is the NN
architecture of the forward map followed by a 2d convolutional neural network.  \delete{A closer
  look at this architecture suggests a further simplification. In fact, the second module of the
  neural network for applying $K^\T$ performs convolution in the $x$ variable, while the third
  module is fully-connected in the $z$ variable. As a result, it is redundant to include an extra
  convolution layer for the deconvolution step afterwards. Due to this consideration, we propose to
  use the same architecture of the forward map for the inverse map.}
The resulting neural network architecture for the inverse map is outlined in \cref{alg:eitinverse}
and illustrated in \cref{fig:eitinverse}. The layers in \cref{alg:eitinverse} share the same
definitions as those in \cref{alg:eitforward} \add{except the $\CNNtwo$ layer, which is defined as
  follows.
\begin{itemize}
\item Post-processing module. $\eta=\CNNtwo[w,n_{\cnn2}](\bar{\eta})$ that maps
  $\bar{\eta}\in\bbR^{N_x\times N_z}$ to $\eta\in\bbR^{N_x\times N_z}$ is a two-dimensional
  convolutional neural network with $n_{\cnn2}$ convolutional layers and $w$ as the window size.
  ReLU is used as the activation function for all intermediate layers. However, as $\eta$ can take
  any real number, the last layer uses a linear activation function.
\end{itemize}
}
\begin{algorithm}[htb]
  \begin{small}
    \begin{center}
      \begin{algorithmic}[1]
        \Require $c, w, n_{\cnn}, n_{\cnn2} \in\bbN$, $\mu\in\bbR^{N_m\times N_h}$
        \Ensure $\eta\in\bbR^{N_x\times N_{{z}}}$
        \State $\tilde{\mu} \leftarrow \encoding[c](\mu)$
        \State $\tilde{\eta} \leftarrow \BCR\first[c, n_{\cnn}](\tilde{\mu})$
        \State $\bar{\eta} \leftarrow \decoding[N_z](\tilde{\eta})$
        \add{\State $\eta \leftarrow \CNNtwo[w, n_{\cnn2}](\bar{\eta})$}
        \State \sffont{return} $\eta$
      \end{algorithmic}
    \end{center}
  \end{small} 
  \caption{\label{alg:eitinverse} Neural network architecture for the one-sided detection inverse
  map $\mu\to \eta$.}
\end{algorithm}

\subsection{Inverse map for the two-sided detection}
For the two-sided detection, the electrodes are placed on both the top and bottom boundaries. The
DtN map hence contains four parts: top-to-top (T2T), top-to-bottom (T2B), bottom-to-top (B2T), and
bottom-to-bottom (B2B). Since the top boundary corresponds to $z=Z$ and the bottom corresponds to
$z=-Z$, the superscripts $+$ and $-$ are used to identify the top and bottom boundaries,
respectively. Following the derivation for the one-sided detection, when $\eta$ is sufficiently
small one can approximate the linearized map from $\eta$ to $\mu^{\pm\pm}$ as
\begin{equation}
  \mu^{\pm\pm}(m, h) \approx 
  \int_{\Omega} \pd{G_0}{n(r)}( (m+h, \pm Z), (x,z))\pd{G_0}{n(s)}( (m-h,\pm Z), (x,z))
  \eta(x,z)\dd x\dd z,
\end{equation}
where the first and second $\pm$ in $\mu^{\pm\pm}$ corresponds to first and second $\pm$ on the
right hand side, respectively. After the discretization, the vector form reads 
\begin{equation}
  \begin{bmatrix}    \mu^{++}\\ \mu^{+-}\\ \mu^{-+}\\ \mu^{--}  \end{bmatrix}
  \approx
  \begin{bmatrix}    K^{++}\\ K^{+-}\\ K^{-+}\\ K^{--}  \end{bmatrix}
  \eta,
  \quad\text{or simply denoted as}\quad
  \mu = K \eta.
\end{equation}

Following the discussion in \cref{sec:forward}, one can factorize each of the four
components $K^{\pm\pm}$ using dimension reduction into three steps.
Hence, the forward map $\eta\to \mu^{\pm\pm}$ can be split into four independent forward
problems for the one-sided detection, and we shall not repeat the study here.

\subsubsection{Architecture for the inverse map}
When $\eta$ is small, the filtered back-projection algorithm for the inverse problem from
$\mu^{\pm\pm}$ to $\eta$ takes the form
\[
  \eta \approx \left( K^\T K +\eps I \right)^{-1}
    \begin{bmatrix}    (K^{++})^\T & (K^{+-})^\T & (K^{-+})^\T & (K^{--})^\T  \end{bmatrix}
  \begin{bmatrix}    \mu^{++}\\ \mu^{+-}\\ \mu^{-+}\\ \mu^{--}  \end{bmatrix}
\]
Following the discussion in \cref{subsec:inv1s}\delete{ and dropping the deconvolution term}, an NN
architecture for the inverse map of the two-sided detection would be to repeat the \add{main part of
  \cref{alg:eitinverse} expect the post-processing module for each of $\mu^{++}, \mu^{+-}, \mu^{-+},
  \mu^{--}$, and then to sum the results together, and to apply the post-processing at last.}

\begin{figure}[htb]
    \centering
    \includegraphics[width=0.95\textwidth]{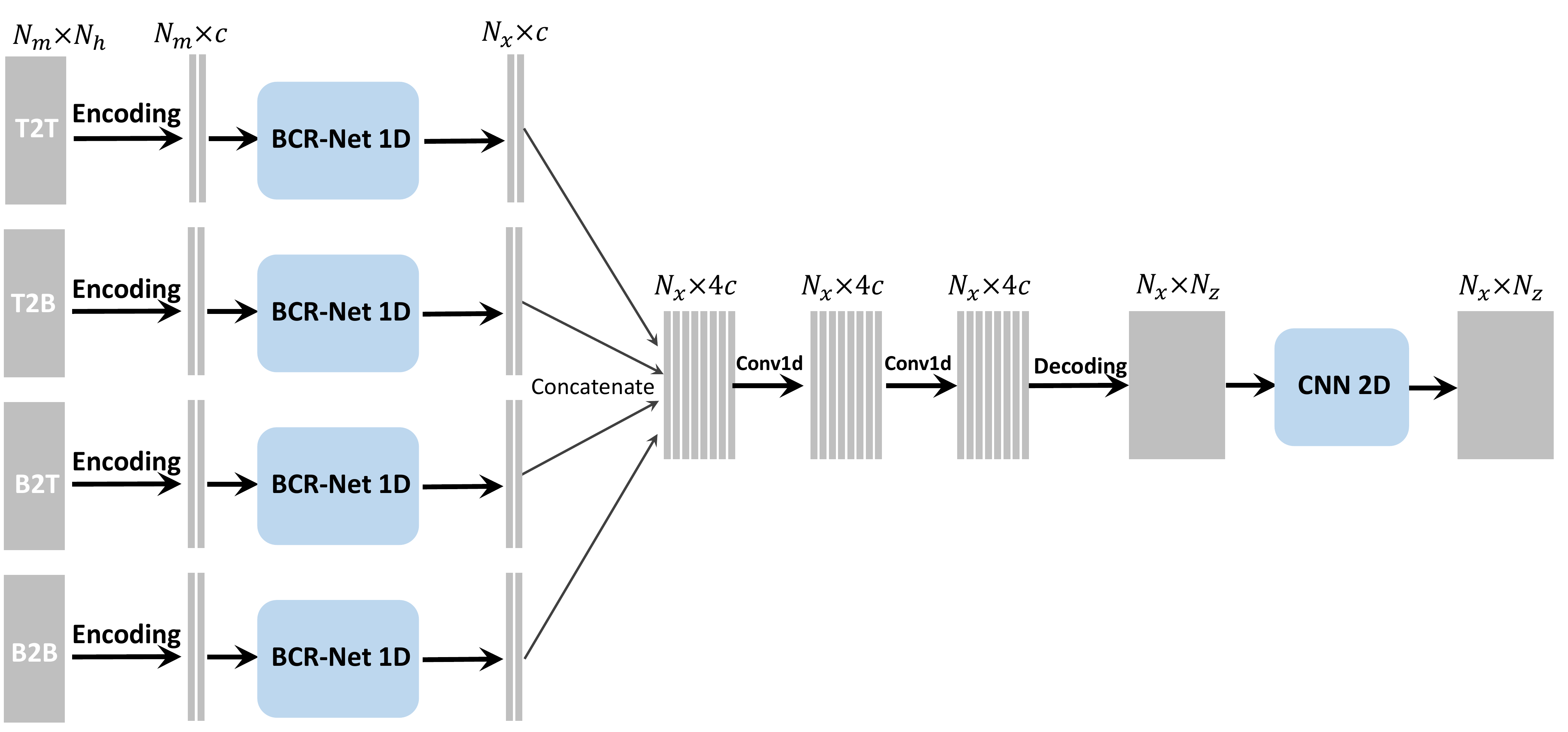}
    \caption{\label{fig:eitTS}Neural network architecture for the inverse problem of the two-sided
    detection.}
\end{figure}

Due to the nonlinearity of the inverse problem, a slightly different approach gives better
performance. Instead of summing the decoded results, one combines, before the decoding step, the
results of $\mu^{++},\mu^{+-},\mu^{-+},\mu^{--}$ into a single array of size $N_x \times 4c$ and
then perform a decoding step together. The resulting neural network architecture is detailed in
\cref{alg:eitTS} and also illustrated in \cref{fig:eitTS}. The modules $\encoding$, $\BCR\first$, 
$\encoding$ and \add{$\CNNtwo$} are same as those in \cref{alg:eitinverse}. The only new layers are
the $\concatenate$ layer: $\eta\leftarrow \concatenate(\eta_1,\eta_2,\eta_3,\eta_4)$, which
concatenates the matrices $\eta_i\in\bbR^{N_x\times c}$, $i=1,2,3,4$ to a matrix with size
$\eta\in\bbR^{N_x\times 4c}$ on the column direction, and the $\Convone[c,w]$ layer: one-dimensional
convolutional layer $\Convone$ with channel number $c$ and window size $w$.

\begin{algorithm}[htb]
  \begin{small}
    \begin{center}
      \begin{algorithmic}[1]
        \Require $c, n_{\cnn}, n_{\cnn2}, n_{\cnn3}, w, w_2\in\bbN$, 
          $\mu^{\pm\pm}\in\bbR^{N_m\times N_h}$ 
        \Ensure $\eta\in\bbR^{N_x\times N_{{z}}}$ 
        \State Denote $\mu_{i}$, $i=1,2,3,4$ by all the cases of $\mu^{\pm\pm}$ 
        \For { $i$ from $1$ to $4$} 
          \State $\tilde{\mu}_i \leftarrow \encoding[c](\mu_i)$
          \State $\tilde{\eta}_i \leftarrow \BCR\first[c, n_{\cnn}](\tilde{\mu}_i)$ 
        \EndFor 
        \State $\tilde{\eta}\leftarrow \concatenate(\tilde{\eta}_1,\tilde{\eta}_2,\tilde{\eta}_3,\tilde{\eta}_4)$ 
        \For {$k$ from $1$ to $n_{\cnn3}$} 
          \State $\tilde{\eta}\leftarrow \Convone[4c, w_2](\tilde{\eta})$ 
        \EndFor 
        \State $\bar{\eta} \leftarrow \decoding[N_x](\tilde{\eta})$ 
        \add{\State $\eta \leftarrow \CNNtwo[w, n_{\cnn2}](\bar{\eta})$}
        \State \sffont{return} $\eta$
      \end{algorithmic}
    \end{center}
  \end{small} 
  \caption{\label{alg:eitTS} Neural network architecture for the two-sided detection inverse map
  $\mu^{\pm\pm}\to \eta$.}
\end{algorithm}

Let us denote $\tilde{\eta}^{\pm,\pm}$ by $\tilde{\eta}_{i}$, $i=1,2,3,4$ in \cref{alg:eitTS},
respectively.  Due to the symmetry of the domain $\Omega$, the map $\mu^{+,+}\to\tilde{\eta}^{+,+}$
is mirror symmetry to the map $\mu^{-,-}\to\tilde{\eta}^{-,-}$. Hence, the B2B part should share the
weights with the T2T part. In the implementation, one can use the same layers for the two maps and
then flip the output for the B2B part to achieve the mirror symmetry. Analogously, the T2B part is
also mirror symmetry to the B2T part. We use the same way to carry out it.

\subsection{Numerical results}\label{sec:numerical2d}

Below we report some numerical results of the neural network proposed above for the 2D EIT
problem. The NN is implemented with Keras \cite{keras} (running on top of TensorFlow
\cite{tensorflow}). Nadam is chosen as the optimizer \cite{dozat2015incorporating} with a step size
$10^{-3}$ and the mean squared error is used as the loss function. The parameters of the network are
initialized randomly from the normal distribution, and the batch size is set as two percent of the
size of the training set. The number of layers in the BCR-Net is set as $n_{\cnn}=6$. \add{For the
  $\CNNtwo$ in \cref{alg:eitinverse,alg:eitTS}, the number of convolutional layers is set as
  $n_{\cnn2}=6$ with window size $w=3$.}  For the one-dimensional convolutional part in
\cref{alg:eitTS}, the number of convolutional layers is set as $n_{\cnn3}=3$ with window size equal
to $w_2=3$. The right value for the channel number $c$ will be studied for each test.

In this section, the half width of the domain $Z$ is set to be $1 / 4$ and the domain $\Omega$ is
discretized by a $160\times 80$ Cartesian grid. \add{Thus, $N_x=N_m=160$, $N_h=40$ and $N_z=80$.}
Both the training data and test data are generated by numerically solving \cref{eq:schrodinger}. In
each test, $10$K pairs of $(\eta,\mu)$ are used to train the neural network and another $10$K pairs
are used as the test data.

For each sample of the training and test data, $\eta$ is randomly sampled and $\mu$ (or
$\mu^{\pm\pm}$) denote the \emph{exact} difference DtN kernel solved by numerical discretization of
\cref{eq:schrodinger}. The predictions of the NNs for the forward and inverse maps are denoted by
$\mu_{NN}$ and $\eta_{NN}$, respectively. The accuracy is measured by the relative error in the
$\ell^2$ norm:
\begin{equation}\label{eq:relativeerror}
  \frac{\|\mu-\mu_{\NN}\|_{\ell^2}}{\|\mu\|_{\ell^2}},\quad
  \frac{\|\eta-\eta_{\NN}\|_{\ell^2}}{\|\eta\|_{\ell^2}}.
\end{equation}
For each experiment, the test error is then obtained by averaging \cref{eq:relativeerror} over a
given set of test samples. The numerical results presented below are obtained by repeating the
training process five times, using different random seeds.

\subsubsection{\add{Smooth potential case.}}\label{sec:continuous}
\add{We first study the smooth potential case, where} the potential $\eta(x)$ is assumed to take the
form
\begin{equation}\label{eq:data}
  \eta(x) = \sum_{i=1}^{n_g} \rho 
  \exp\left( -\frac{1}{2}(x-c^{(i)})^\T(\Theta^{(i)})^{-1}(x-c^{(i)}) \right),
\end{equation}
with $\rho=1000$. Each matrix $\Theta^{(i)}\in\bbR^{2\times 2}$ is generated with the eigenvalues
uniformly sampled in $[0.0125, 0.05]$ and the eigenvectors uniformly sampled in the unit circle
$\bbS^1$. Two types of data sets are generated to test the neural networks.
\begin{itemize}
\item Shallow inclusions. The centers of Gaussians are sampled as
  $c^{(i)}\in\cU([0,1]\times[0.05,0.2])$. This data is used to test the forward and inverse problem
  for the one-sided detection.
\item Deep inclusions. The centers of Gaussians are sampled as
  $c^{(i)}\in\cU([0,1]\times[-0.2,0.2])$. This data can be used to show the instability of the
  inverse map: the one-sided detection would fail to resolve the inverse problem well, while the
  two-sided detection works.
\end{itemize} 

\paragraph{One-sided detection for shallow inclusions.}

\begin{figure}[h!]
  \centering
  \subfloat[One-sided detection forward problem]{
    \includegraphics[width=0.35\textwidth,clip]{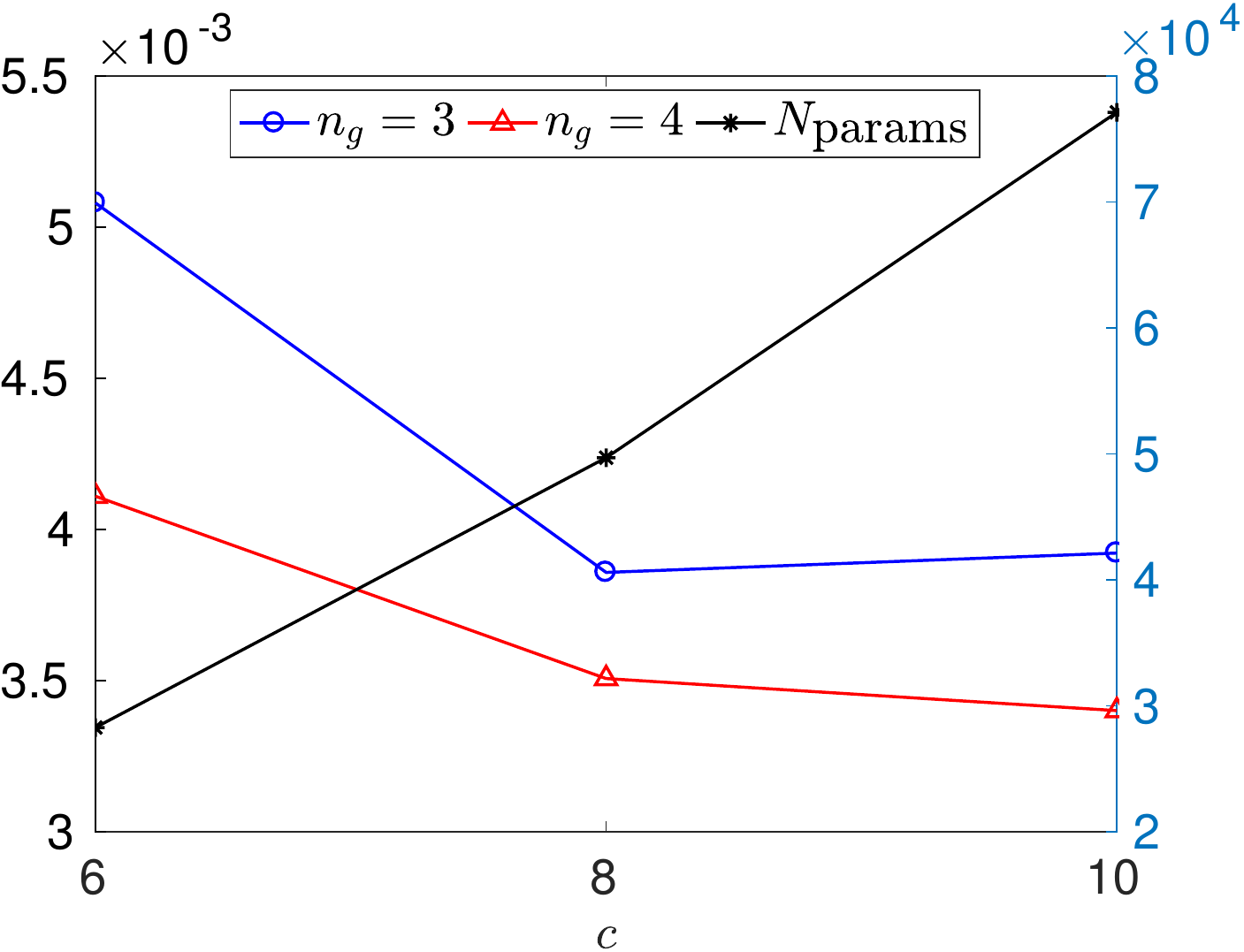}
  }\hspace{0.1\textwidth}
  \subfloat[One-sided detection inverse problem]{
    \includegraphics[width=0.35\textwidth,clip]{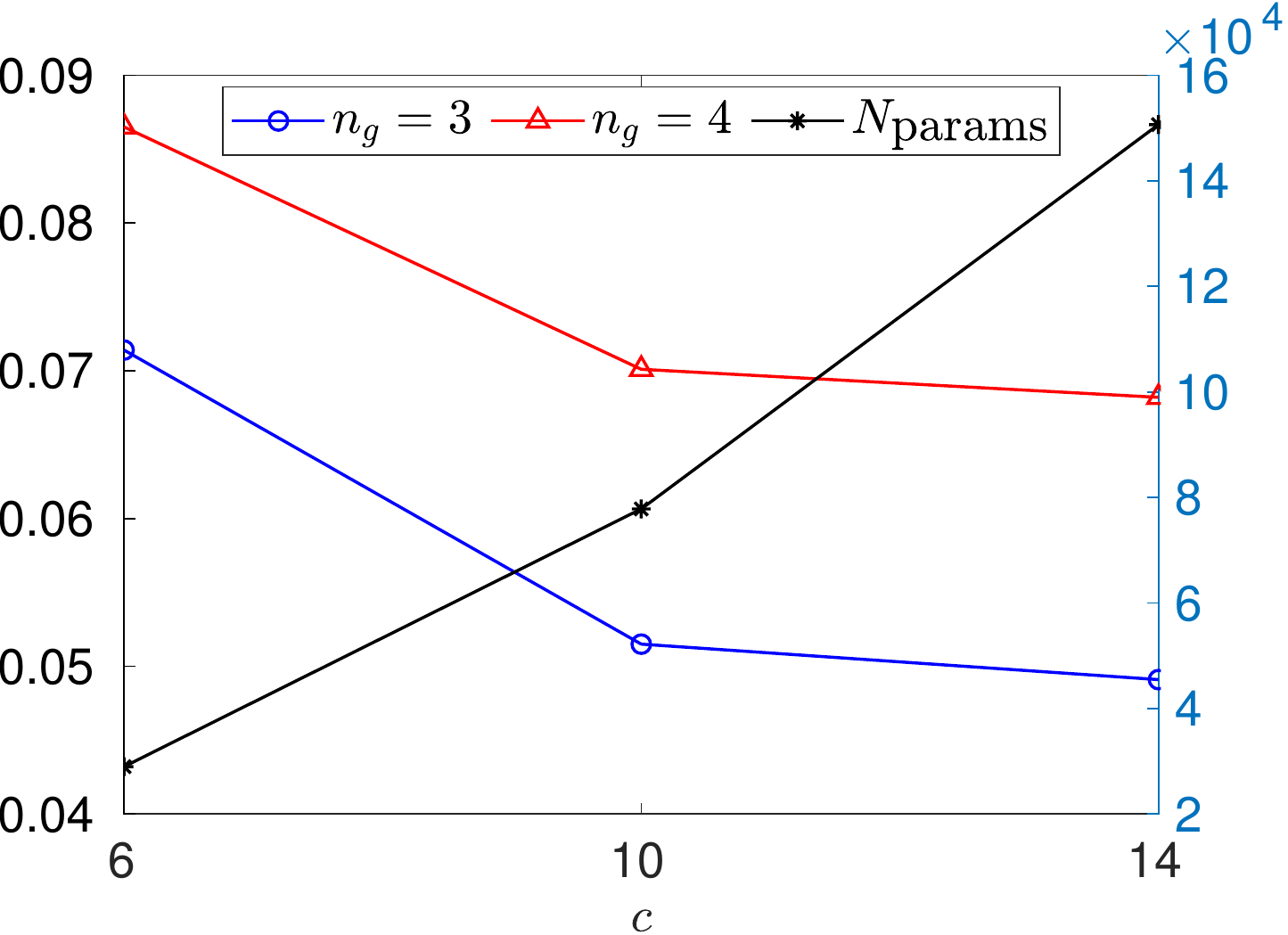}
  }
  \caption{\label{fig:1sShallow}The relative error and the number of parameters for the one-sided
  detection forward and inverse problem for different number of channels $c$ and different number of
  Gaussians $n_g$ with shallow inclusions.}
\end{figure}

\begin{figure}[h!]
  \centering
  \begin{minipage}{0.45\textwidth}
    \subfloat[$\mu$]{
      \includegraphics[width=\textwidth]{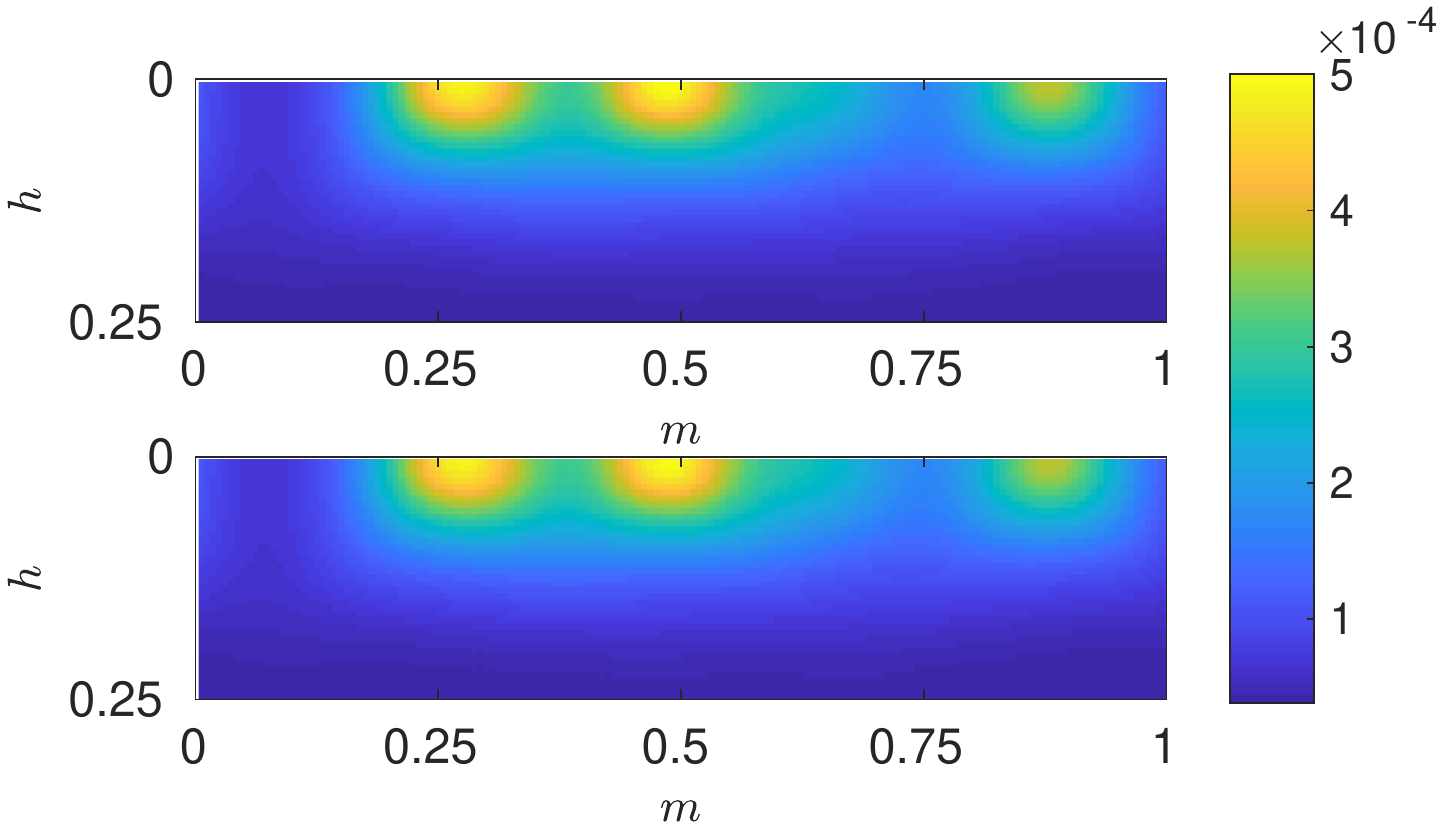}
    }\\
              \end{minipage}\hspace{0.04\textwidth}
  \begin{minipage}{0.45\textwidth}
    \subfloat[$\eta$]{
      \includegraphics[width=\textwidth]{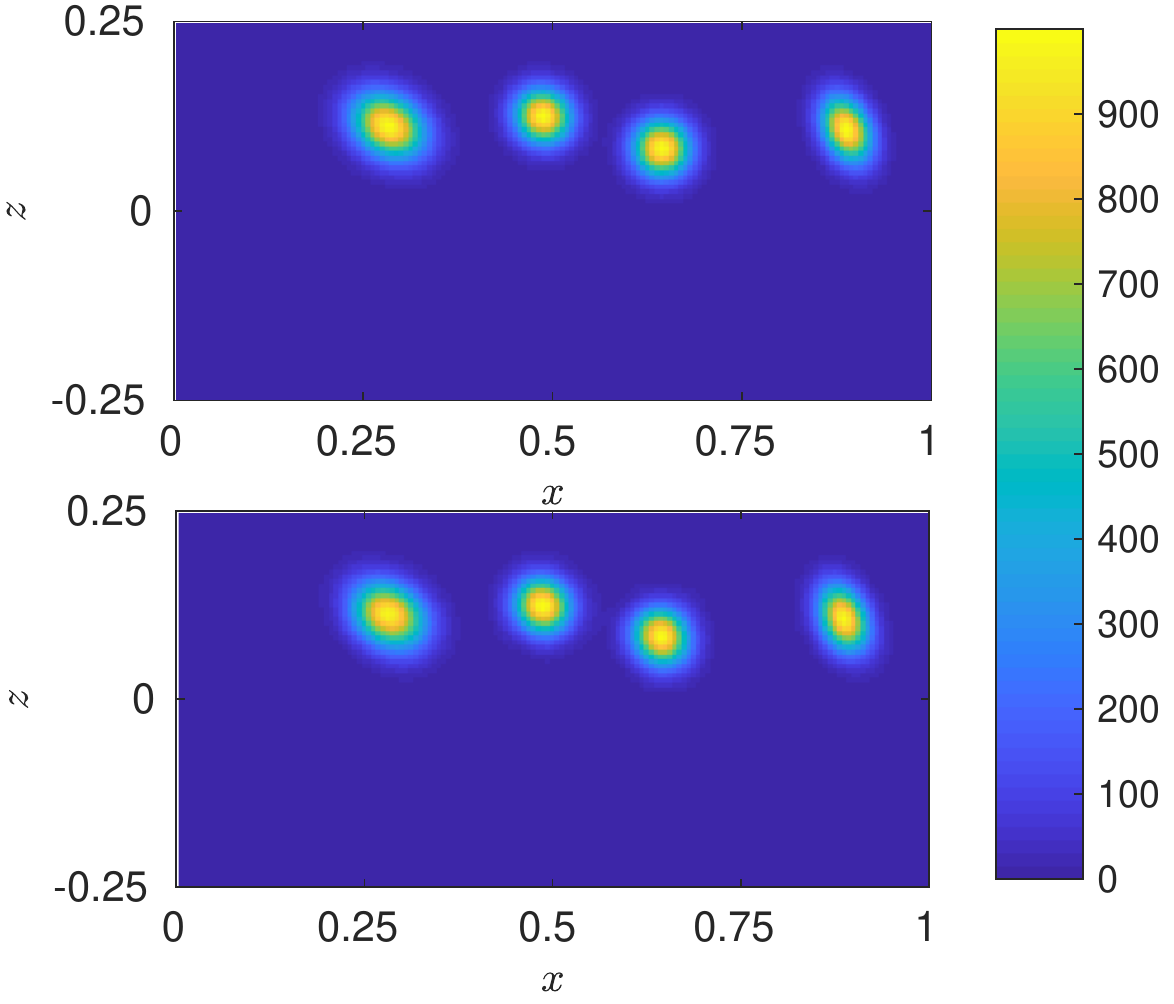}
    }
  \end{minipage}
  \caption{\label{fig:1sShallowFig} A sample in the test data for the one-sided detection 
    with shallow inclusions with $n_g=4$.
        (a) The data $\mu$;     (b) The potential $\eta$.
    In each figure, the upper one is the reference solution and the lower one is the prediction of
    the neural network. For the forward problem, the channel number of the neural network is 
    $c=8$ and the relative error for this sample is $3.9e-3$. For the inverse problem, 
    the channel number is \revised{$c=9$}{$c=10$} and the relative error for this sample is 
  \revised{$6.5e-2$}{$6.3e-2$}.}
\end{figure}

\Cref{fig:1sShallow} gives the test error and the number of parameters for both the forward and
inverse maps with different values of $c$ (the channel number) and $n_g$ (the number of
Gaussians). The NN predictions $\mu_{NN}$ and $\eta_{NN}$ along with the exact solutions are
illustrated in \cref{fig:1sShallowFig}.  For the forward problem, the test error is relatively small
even for $c=6$, where only $28$K parameters are used in the neural network (compared with the size
$12,800=160\times 80$ of $\eta$). As the number of channels $c$ increases, the test error decays
first and then stagnates. The choice of $c=8$ is a balance between accuracy and efficiency for this
forward problem.


For the inverse problem, the test error is relatively small when \revised{$c=9$}{$c=10$}, where
\revised{$63$K}{$78$K} parameters are used in the neural network. Judging from the plots, the neural
network produces accurate results in term of the location, the shape, and the magnitude of the
Gaussians. These results demonstrate that the neural network architectures proposed in this section
are capable of representing the forward and inverse maps for shallow inclusions.

\begin{figure}[h!]
    \centering
    \subfloat[One-sided detection forward problem]{
    \includegraphics[width=0.35\textwidth,clip]{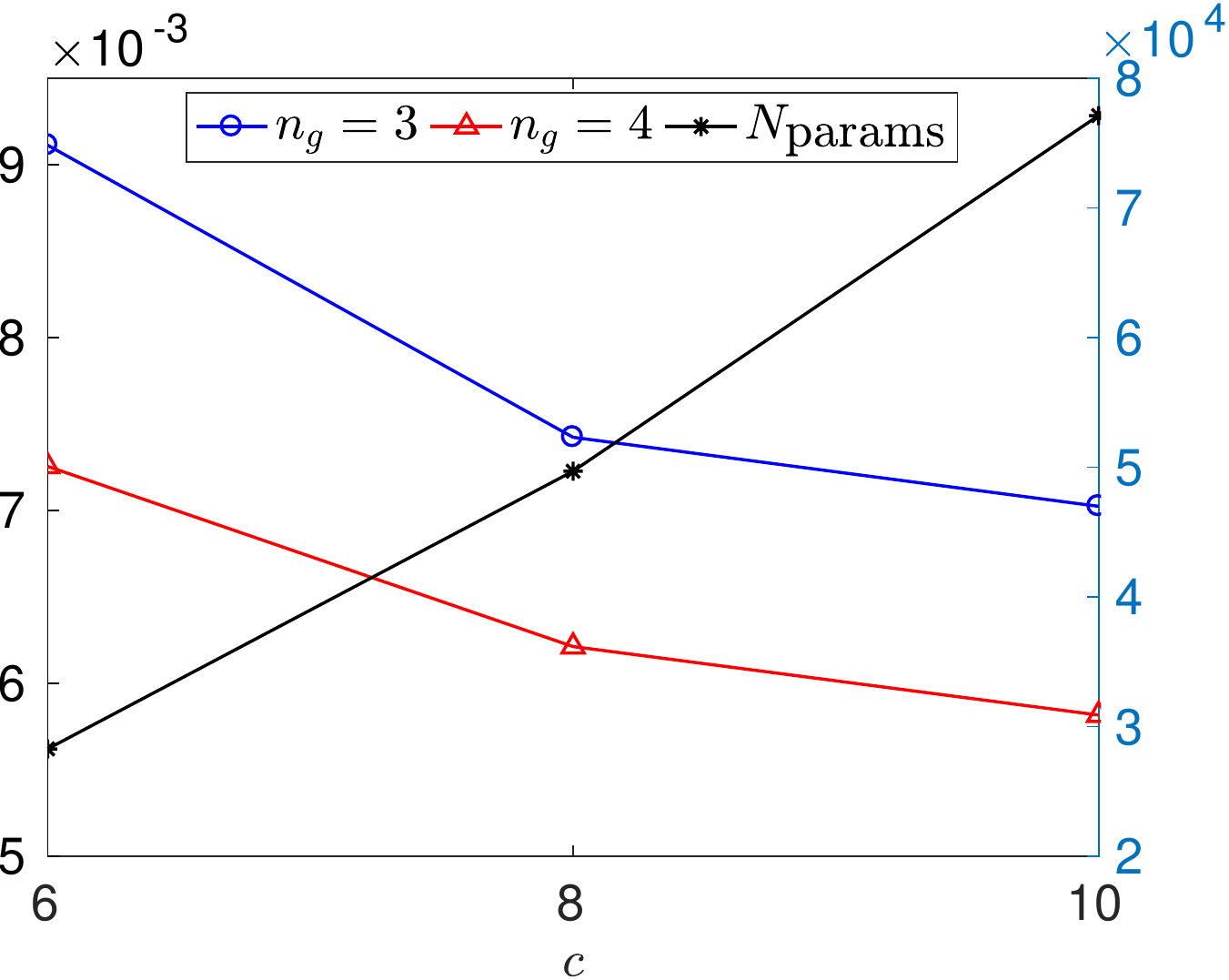}
    }\hspace{0.1\textwidth}
    \subfloat[One-sided detection inverse problem]{
    \includegraphics[width=0.35\textwidth,clip]{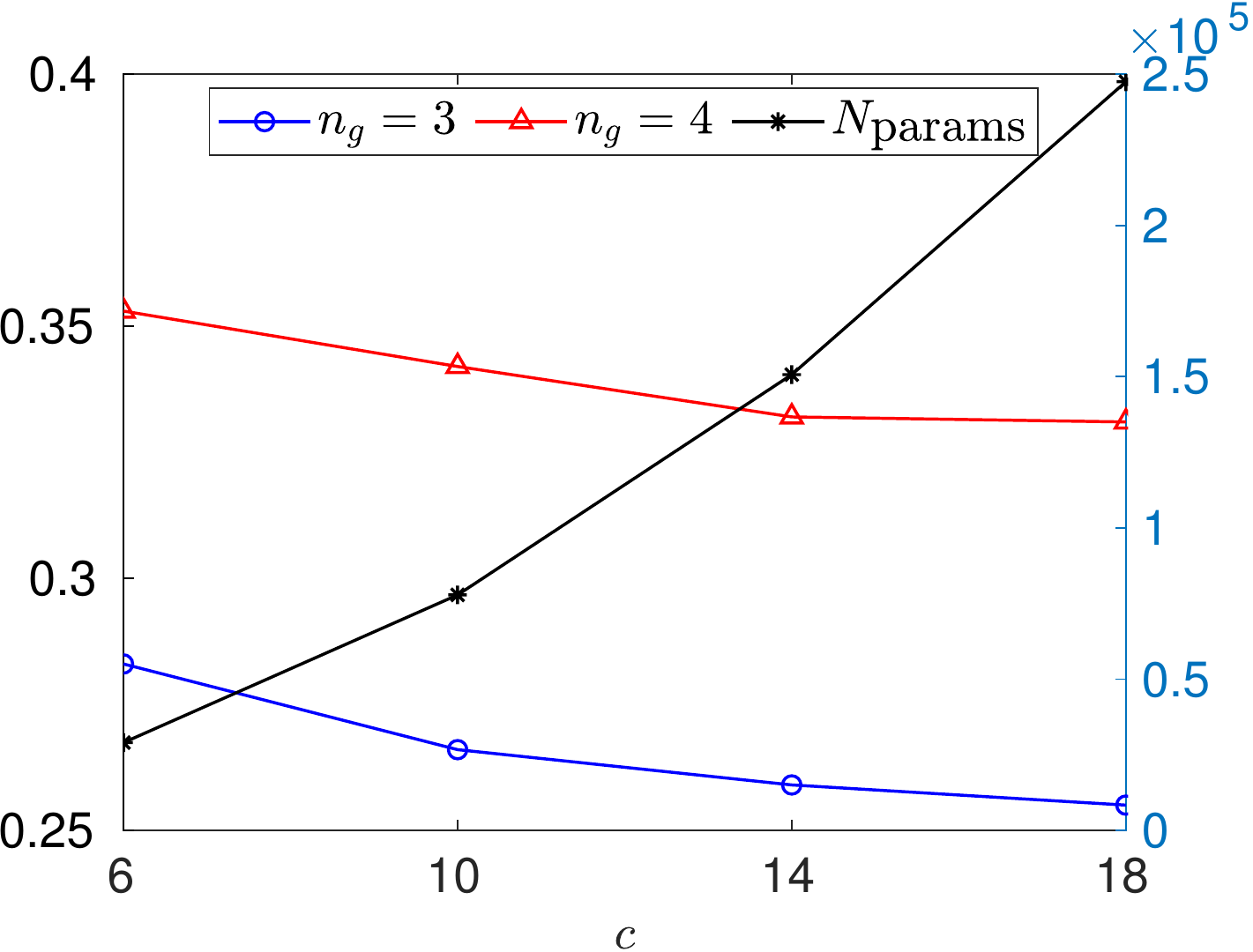}
    }
    \caption{\label{fig:1sDeep}The relative error and the number of parameters for the one-sided
      detection forward and inverse problem for different values of $c$ and $n_g$ with deep
      inclusions.}
\end{figure}

\begin{figure}[h!]
    \centering
    \begin{minipage}{0.45\textwidth}
        \subfloat[$\mu$]{
        \includegraphics[width=\textwidth]{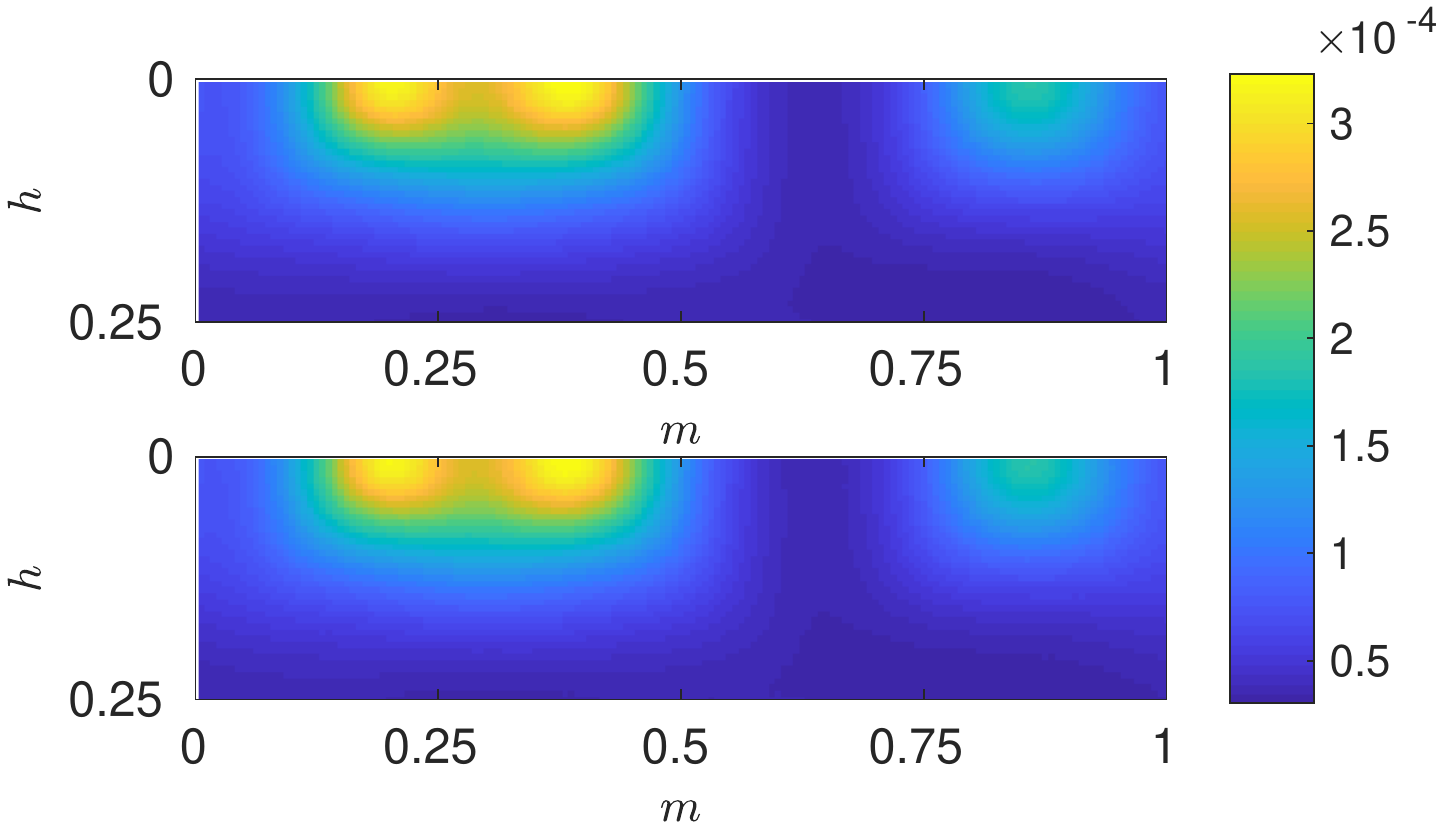}
        }\\
                            \end{minipage}\hspace{0.04\textwidth}
    \begin{minipage}{0.45\textwidth}
        \subfloat[$\eta$]{
        \includegraphics[width=\textwidth]{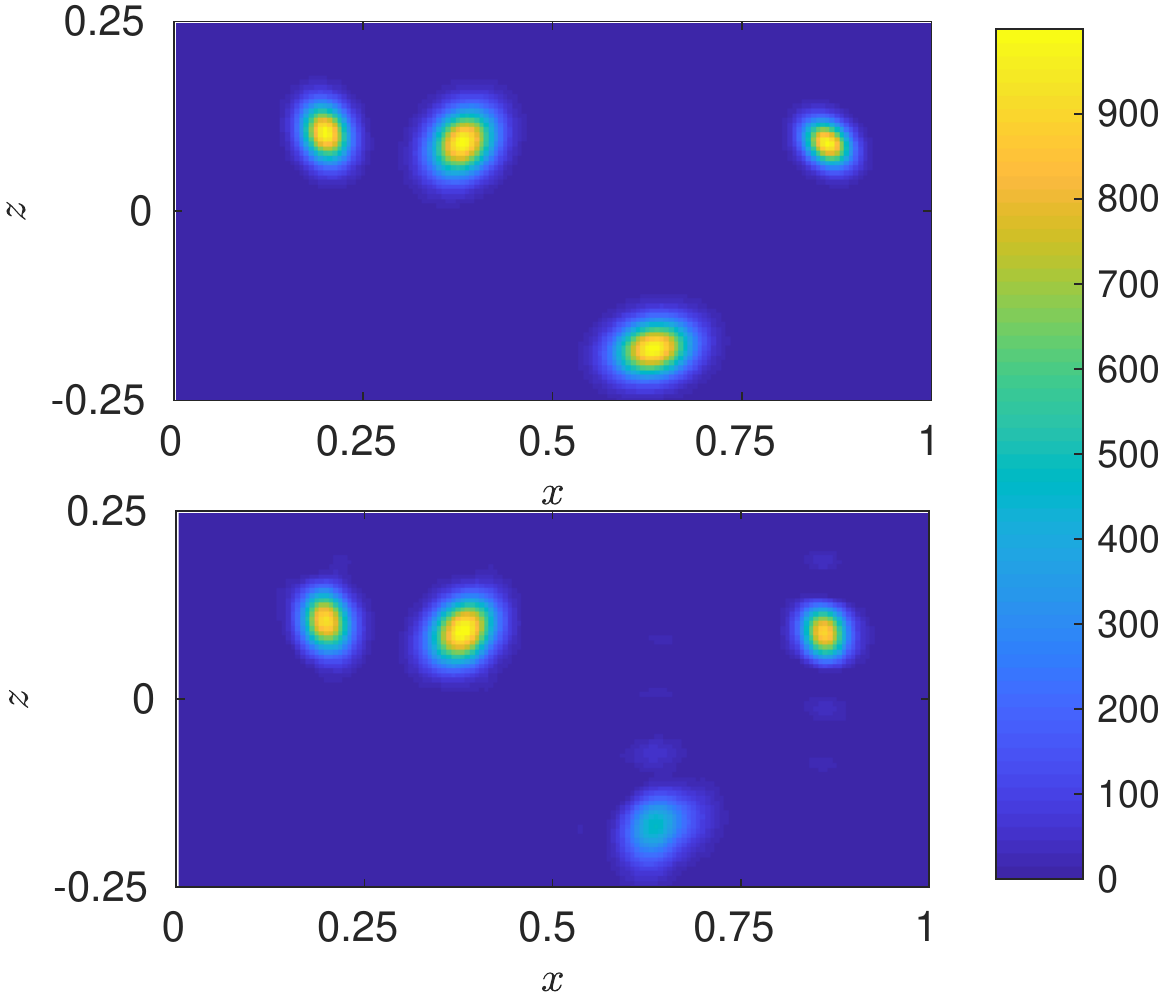}
        }
    \end{minipage}
    \caption{\label{fig:1sDeepFig} A sample in the test data for the one-sided detection with deep
    inclusions.
            (a) The data $\mu$; (b) The
      potential $\eta$.  In each figure, the upper one is the reference solution and the lower one
      is the prediction of the neural network. For the forward problem, the channel number of the
      neural network is $c=8$ and the relative error for this sample is $5.8e-3$. For the inverse
      problem, the channel number is \revised{$c=12$}{$c=14$} and the relative error for this sample
    is \revised{$0.34$}{$0.30$}.}
\end{figure}

\paragraph{One-sided detection for deep inclusions.}
\cref{fig:1sDeep} plots the test error and number of parameters for different values of $c$ and
$n_g$. The predicted $\mu_{NN}$ and $\eta_{NN}$ and the reference solution of one specific test
sample are presented in \cref{fig:1sDeepFig}. For the forward map, the test error is comparable with
the case of shallow inclusions. However, the neural network for the inverse map fails to produce a
good prediction.  In \cref{fig:1sDeepFig}, the prediction is pretty close near the top boundary but
gives significant error near the bottom boundary. This result agrees with the conclusion in
\cite{allers1991stability} that the resolution near the boundary is much better than deep in the
interior, which is caused by the instability of the inverse problem
\cite{alessandrini1997examples,allers1991stability,uhlmann2009electrical}. To avoid it, more
information on the object must be provided, for instance, add electrodes on the bottom boundary,
i.e., the two-sided detection.


\paragraph{Two-sided detection for deep inclusions.}
As we have seen, due to the instability of the inverse problem, the one-sided detection fails to
resolve the problem with deep inclusions. Here we test the neural network for the two-sided
detection for deep inclusions.

\begin{figure}[h!]
    \centering
        \includegraphics[width=0.35\textwidth,clip]{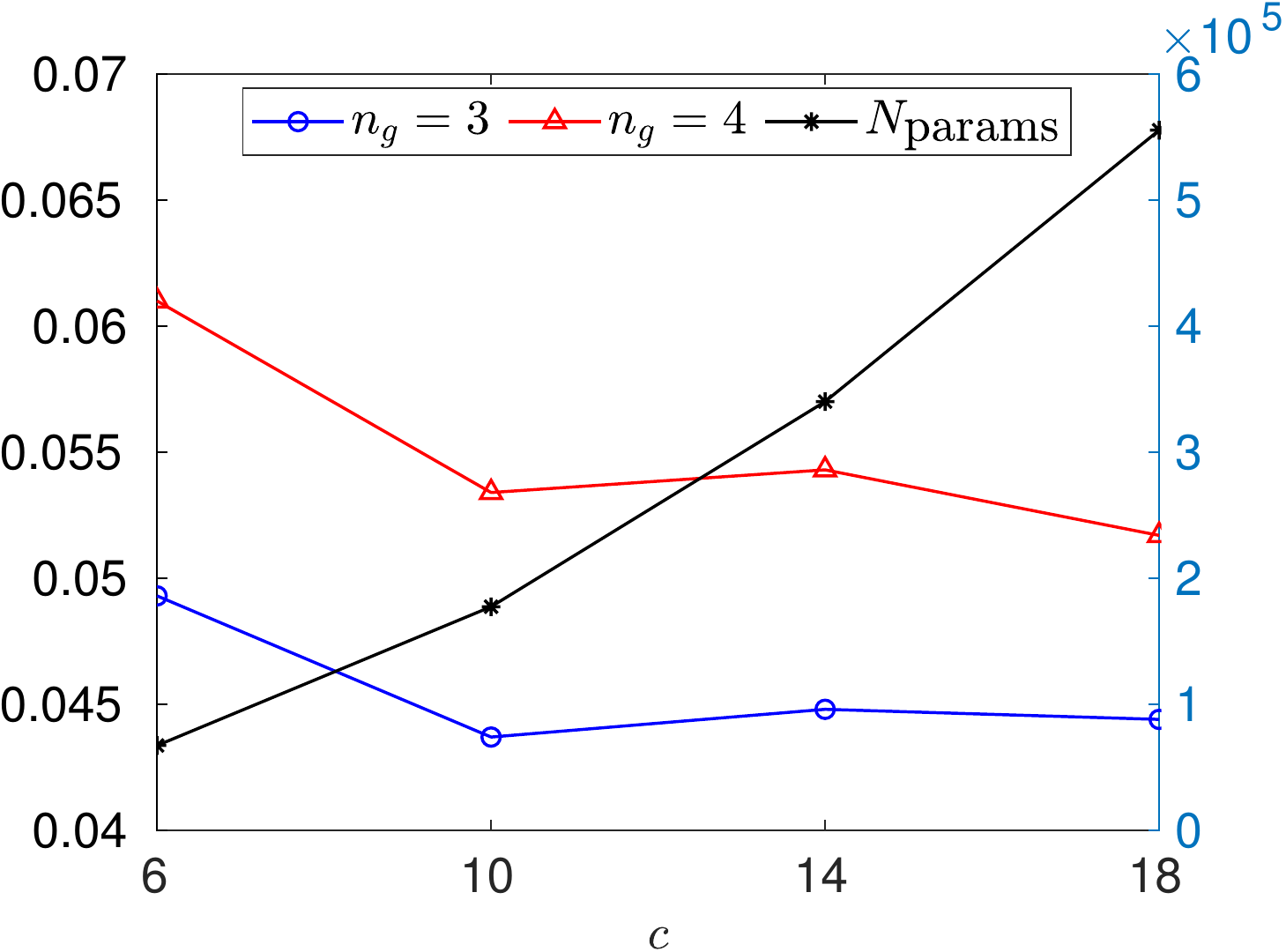}
        \caption{\label{fig:TSerr}The relative error and the number of parameters for the
    two-sided detection inverse problem for different number of channels $c$ and different number of
  Gaussians $n_g$ with deep inclusions.}
\end{figure}

\begin{figure}[h!]
    \centering
    \subfloat[$\mu^{\pm\pm}$]{
    \includegraphics[width=0.38\textwidth]{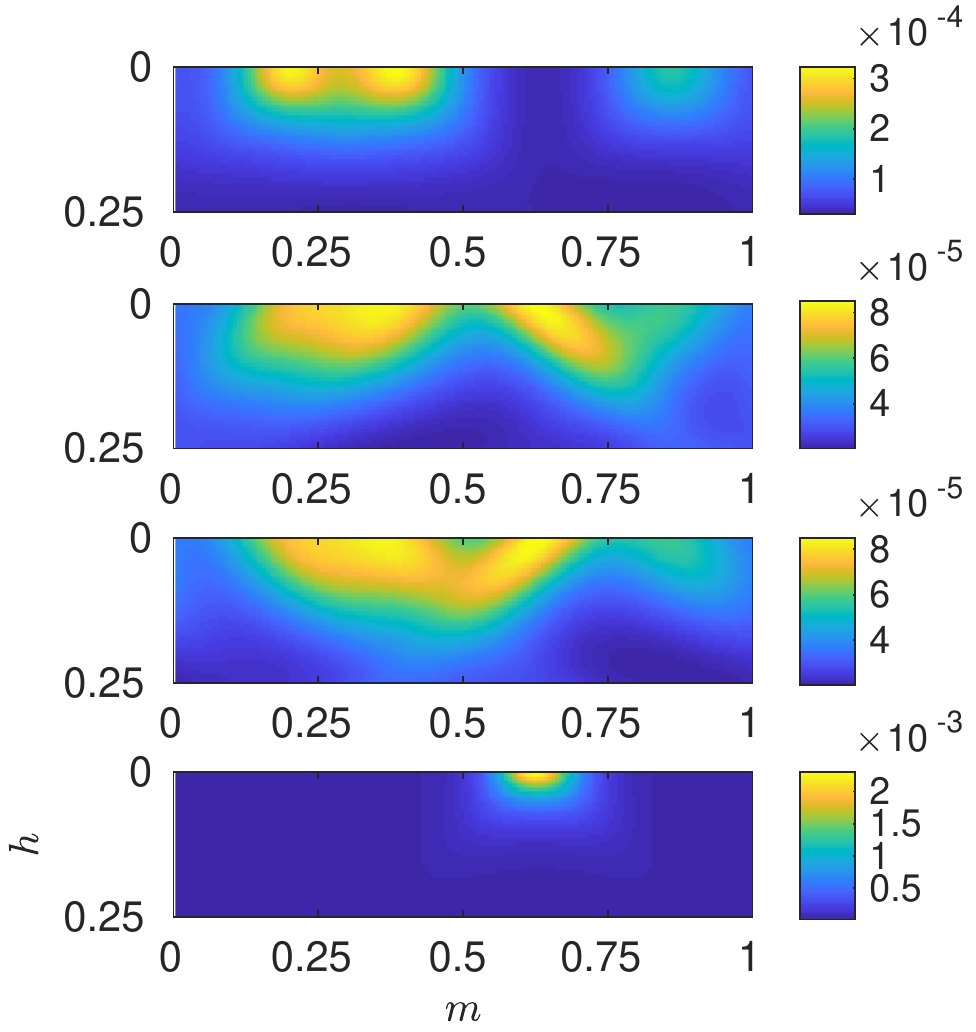}
    }\hspace{0.1\textwidth}
    \subfloat[$\eta$]{
    \includegraphics[width=0.45\textwidth]{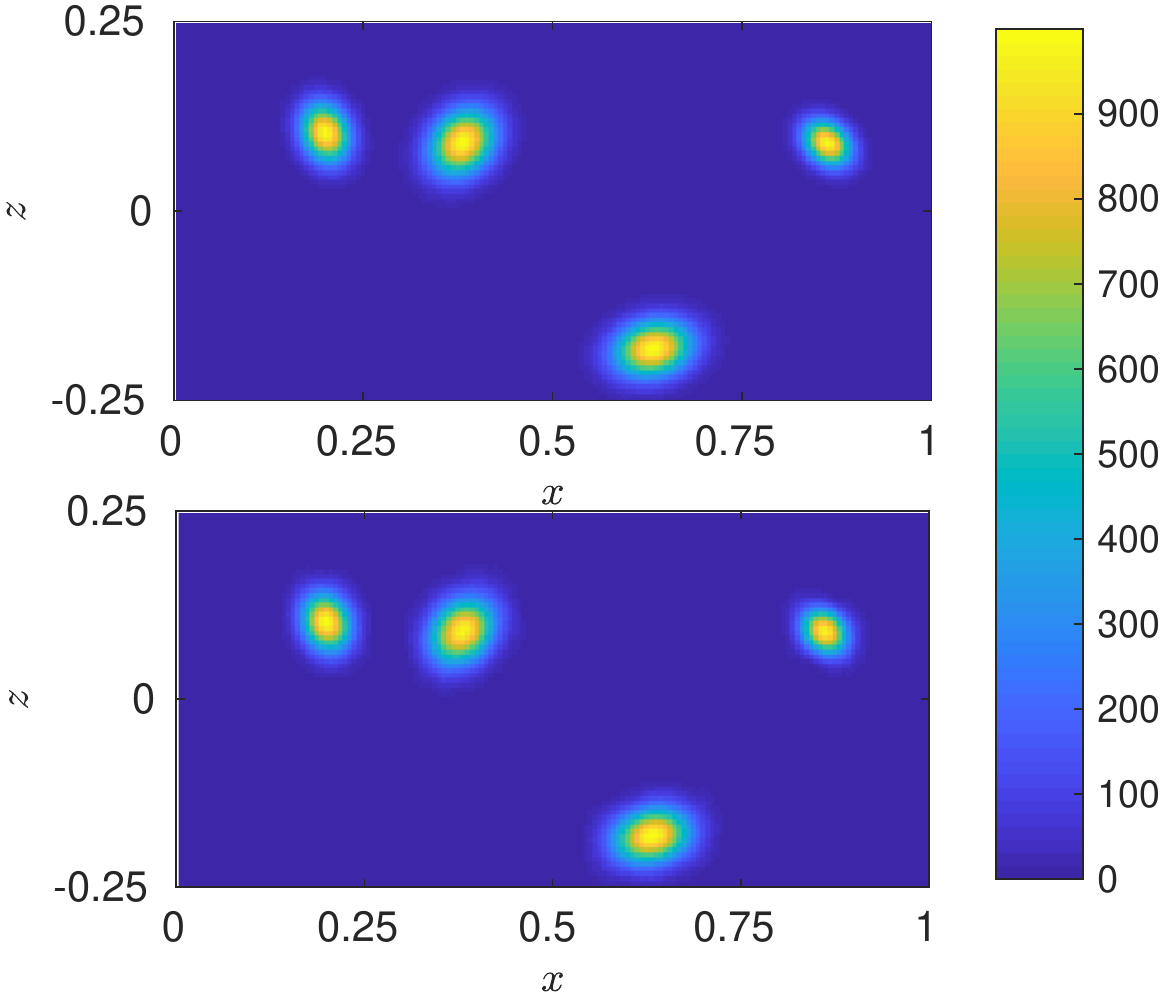}
    }
    \caption{\label{fig:TSfig} A sample in the test data for the two-sided detection inverse problem
      with \revised{$c=9$}{$c=10$} and $n_g=4$. (a) The data $\mu^{\pm\pm}$; (b) The exact solution
    $\eta$ (the upper one) and the prediction $\eta_{NN}$ (the lower one). The relative error for
  this sample is \revised{$3.5e-2$}{$3.1e-2$}.}
\end{figure}

\Cref{fig:TSerr} shows the test error and number of parameters for different values of $c$ and
$n_g$.  The NN predictions $\mu_{NN}$ and $\eta_{NN}$ along with the reference solution of the same
sample in \cref{fig:1sDeepFig} are summarized in \cref{fig:TSfig}. The test error is significantly
decreased and is even slightly less than that in \cref{fig:1sShallow} of the one-sided problem for
the shallow inclusions. Notice that the test error is relatively small even for the case
\revised{$c=9$}{$c=10$} with \revised{$140$K}{$177$K} parameters in the neural network.

\add{ 
\subsubsection{Shape reconstruction}
\cref{sec:continuous} studied the behavior of the proposed NN architectures for smooth potentials
and show that the proposed NN architectures have the ability to give a good prediction on both the
forward and inverse problems. Here the focus is on shape reconstruction with noise measurement. The
numerical results in \cref{fig:1sShallow,fig:TSerr} show that the choice of channel number $c=10$ is
a good balance between accuracy and efficiency for the inverse problem. To simplify the discussion,
we set $c=10$ in all the test in this subsection.

The potential $\eta(x)$ is assumed to be a piecewise constant. Four shapes are placed in $\Omega$,
where each can be a regular triangle, square, pentagon and hexagon. $\eta(x)$ is set to $1000$ in
the shapes and is $0$ otherwise.  For each shape, the circumcircle radius is uniformly sampled in $[0.05,0.1]$
and the direction is uniformly sampled in the unit circle $\bbS^1$. For the shallow inclusion case,
the center the shape is uniformly sampled in $[0,1]\times [0.05, 0.2]$, while it is uniform from
$[0,1]\times [-0.2,0.2]$ for the deep inclusion case.

To model the uncertainty in the measurement data, noises have been added to the DtN map in the data
set by setting $\lambda_{\eta, i}^{\delta}\equiv (1+Z_i\delta)\lambda_{\eta,i}$, where $Z_i$ is a
Gaussian random variable with zero mean and unit variance and $\delta$ controls the signal-to-noise
ratio. In the following tests, the noise level is chosen as $\delta=0, 0.5\%$ and $1\%$. In the
numerical experiments, an independent NN is trained and tested with the noise data set
$\{\mu_i^\delta, \eta\}$ with $\mu_i^\delta\equiv \lambda_{\eta,i}^\delta-\lambda_0$ for each noise
level.  It is worth noting that the mean of $\frac{\|\lambda_{\eta}-\lambda_0\|}{\|\lambda_{\eta}\|}$
for all the samples for the shallow or deep inclusions cases are both about $10\%$ and hence the
signal-to-noise ratio for the difference $\mu$ is about $10\cdot\delta$.

\paragraph{One-sided detection for shallow inclusions}

\Cref{fig:shape_shallow_ng4} shows a sample in the test data for different noise level $\delta$.
When there is no noise in the measurement data, the NN provides a good prediction of the potential
$\eta$ in terms of both shape and position. As is noted in \cref{sec:continuous}, since the inverse
problem is ill-posed, the resolution near the boundary is accurate while the shape boundaries deep
in the interior is blurry. For the same reason, when there is noise in the measurement data, the
shape boundaries become more blurry as the noise level grows. However, the positions and number of
shapes are correctly predicted.

\begin{figure}[h!]
  \centering
  \includegraphics[width=0.8\textwidth]{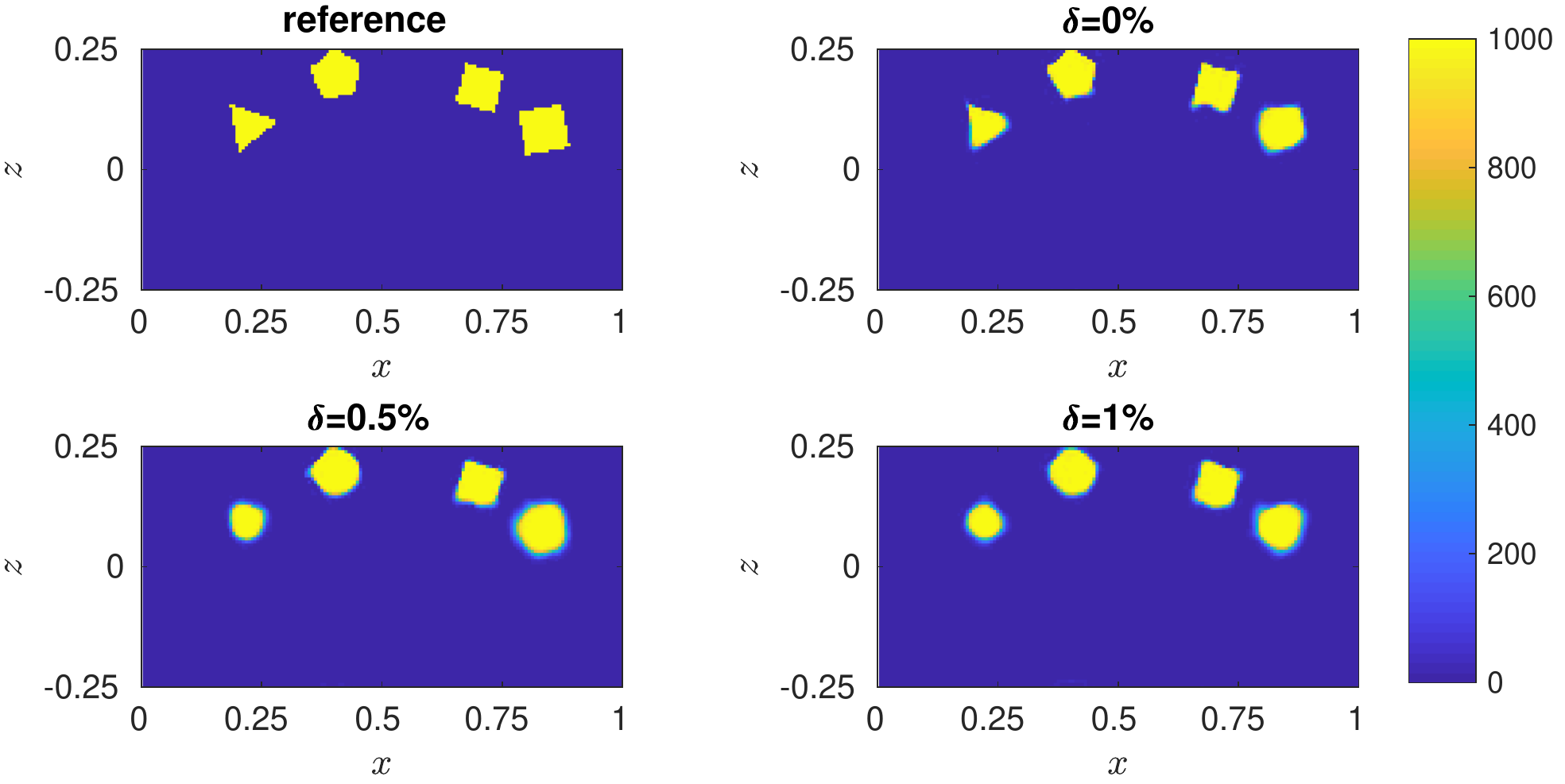} 
  \caption{\label{fig:shape_shallow_ng4}A sample in the test data for the one-sided detection for the
    shape reconstruction with shallow inclusion for different noise level $\delta=0\%,
  0.5\%, 1\%$.}
\end{figure}

\paragraph{Two-sided detection for deep inclusions}
\Cref{fig:shape_deep_ng4} presents a sample from the test data for different noise levels $\delta$
for the deep inclusion. The conclusions for the one-sided detection for shallow inclusions still
hold for this case. Comparing the results in \cref{fig:shape_shallow_ng4,fig:shape_deep_ng4}, one
finds that the NN for the two-sided detection gives a better prediction for the shapes in the middle
(i.e. close to $z=0$). This agrees with the fact that the two-sided detection utilizes more
information (not only $\mu^{++}$ and $\mu^{--}$, but also $\mu^{+-}$ and $\mu^{-+}$).

\begin{figure}[h!]
  \centering
  \includegraphics[width=0.8\textwidth]{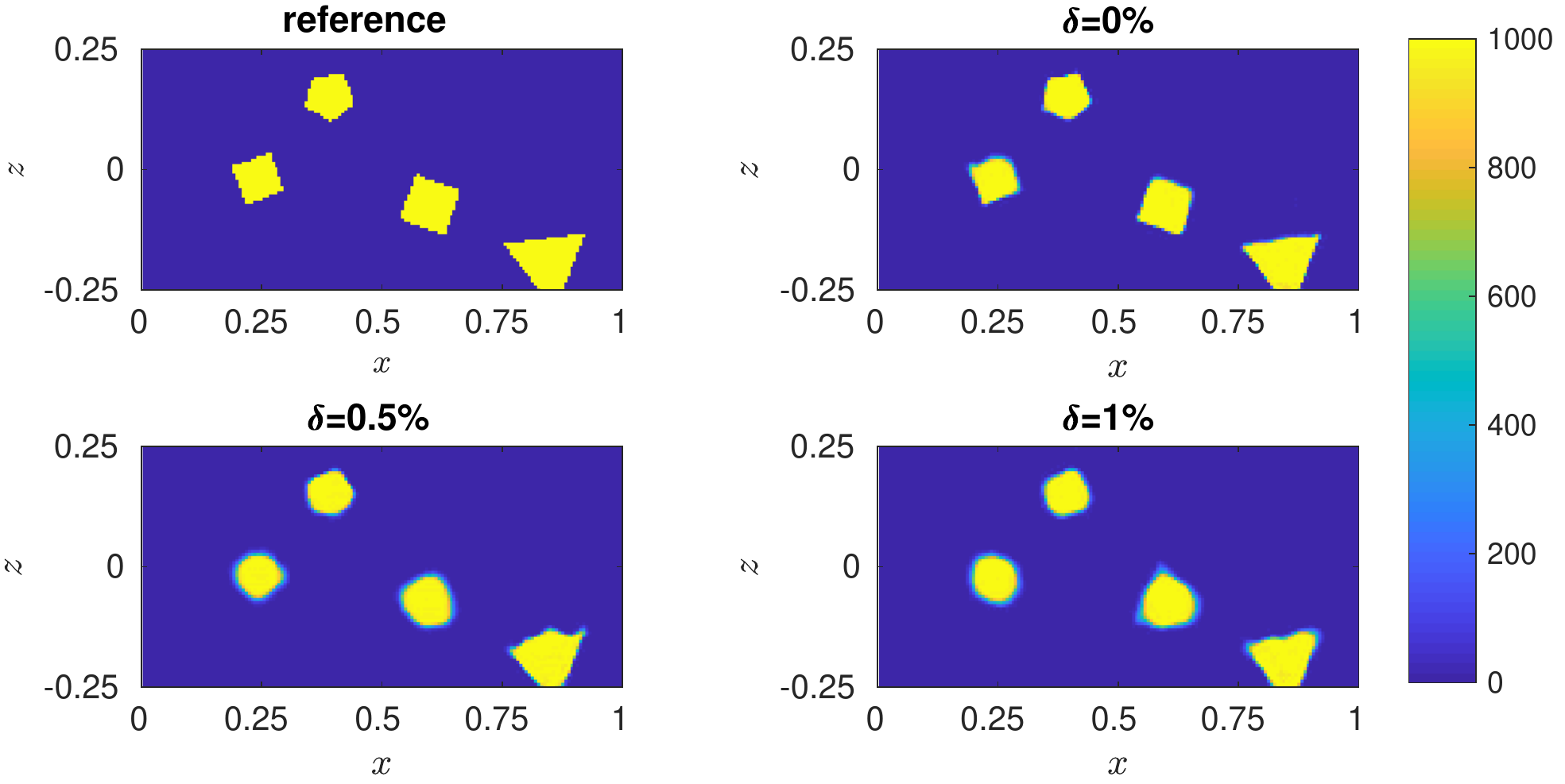} 
  \caption{\label{fig:shape_deep_ng4}A sample in the test data for the two-sided detection for the
    shape reconstruction with deep inclusion with different noise level $\delta=0\%,
  0.5\%, 1\%$.}
\end{figure}

All the numerical tests show that the NNs in \cref{alg:eitforward,alg:eitinverse,alg:eitTS} are
capable of learning the forward and inverse problem of EIT for various setups.  }

\section{Neural network for 3D the case}\label{sec:3d}

\begin{figure}[htb]
  \centering
  \subfloat[One-sided detection]{
    \includegraphics[width=0.49\textwidth,page=1]{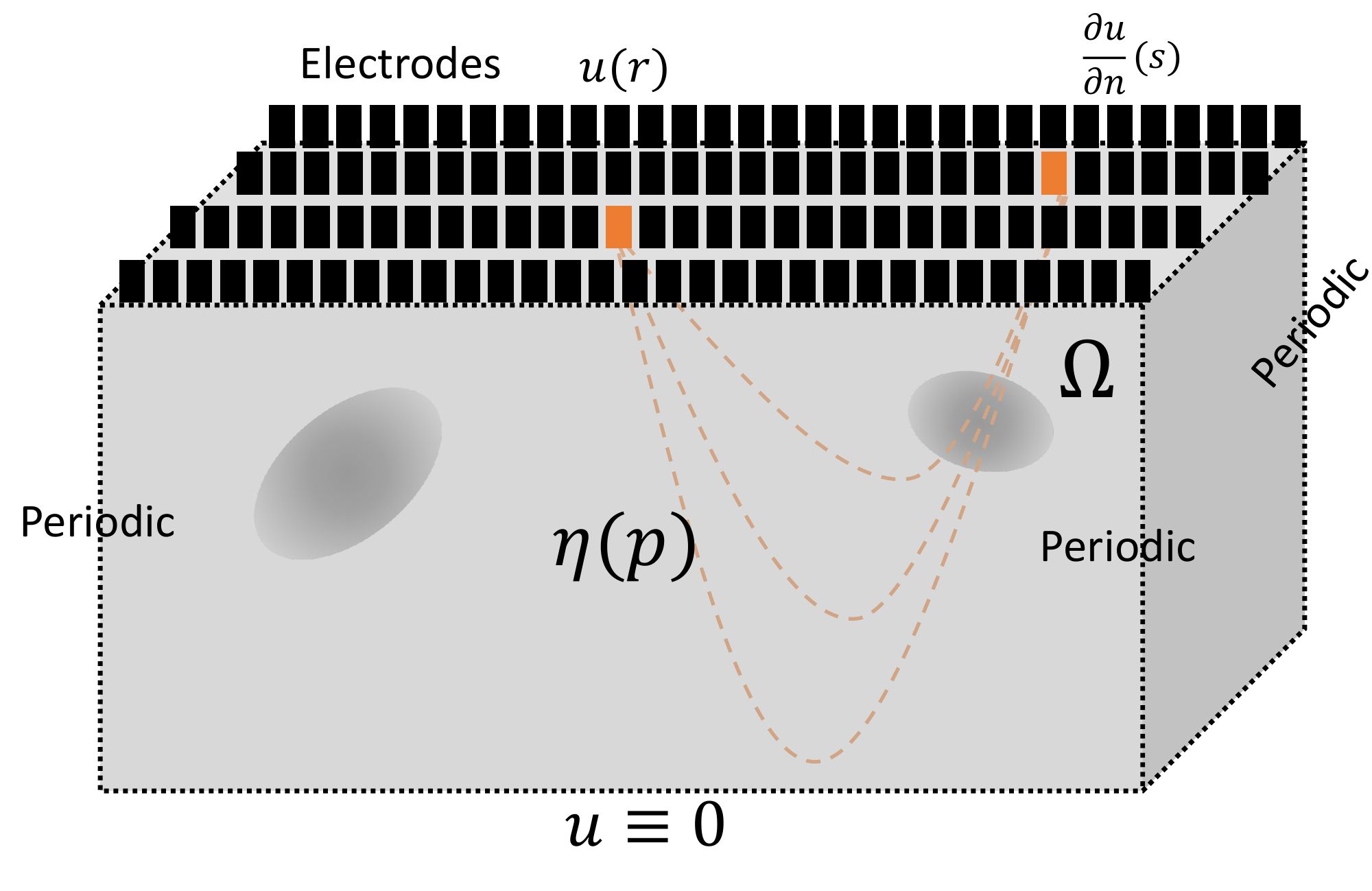}
  }
  \subfloat[Two-sided detection]{
    \includegraphics[width=0.49\textwidth,page=2]{domain3.pdf}
  }
  \caption{\label{fig:domain3}Illustration of the problem setup for 3D case. In the one-sided
    detection, the electrodes are only placed on the top and the zero Dirichlet boundary condition
    is applied on the bottom. In the two-sided detection, the electrodes are placed on both the top
    and bottom. In both cases, periodic boundary conditions are applied on the left and right, and
    front and back boundaries.}
\end{figure}

For the 3D case, the domain is assumed to be $\Omega=[0,1] \times [0,1] \times [-Z,Z]$. The periodic
boundary condition is applied on the left, right, front, and back for simplicity.  Similar to the 2D
case, the electrodes are allowed to be placed on either the top (one-sided detection) or both top
and bottom (two-sided detection), as shown in \cref{fig:domain3}. For the one-sided detection, the
zero Dirichlet boundary condition is applied on the bottom boundary.

\subsection{Analysis and NN architecture}\label{sec:analysis3d}

Most of the 2D analysis for the maps $\eta\to\mu$ and $\mu\to\eta$ can be extended to the 3D
case. The main difference is that the data $\mu$ is a four-dimensional function while the potential
$\eta$ is a three-dimensional function. Below we will first study the extension for the one-sided
detection and then briefly discuss the two-sided detection.

\paragraph{One-sided detection.}
The DtN map for the one-sided detection is limited on the top boundary. Let $r=(r_1,r_2,Z)$,
$s=(s_1,s_2,Z)$, and $p=(x,y,z)$, where $x,y$ are for the horizontal directions and $z$ is for the
depth direction. The map \cref{eq:V2lambda} for the 3D case can be written as
\begin{equation}\label{eq:approx3d}
  \mu( (r_1, r_2, Z), (s_1,s_2,Z)) \approx
  \int_{\Omega}\pd{G_0}{n(r)}( (r_1, r_2,Z), (x,y, z))
  \pd{G_0}{n(s)}( (s_1,s_2, Z),(x,y, z))\eta(x,y,z)\dd x\dd y\dd z.
\end{equation}
Introducing the new variables $m=(m_1,m_2)$ and $h=(h_1,h_2)$ with $r_1=m_1+h_1$, $r_2=m_2+h_2$,
$s_1=m_1-h_1$ and $s_2=m_2-h_2$ yields
\begin{equation}\label{eq:def_kappa3d}
  \mu(m,h) :=\mu( (m_1+h_1, m_2+h_2,Z), (m_1-h_1,m_2-h_2,Z) ) \approx
  \int_{\Omega}K(m,h,x,y,z)\eta(x,y,z)\dd x\dd y\dd z,
\end{equation}
where 
\begin{equation}\label{eq:def_K3d}
    K(m,h,x,y,z) := \pd{G_0}{n}( (m_1+h_1,m_2+h_2, Z), (x,y, z)) 
    \pd{G_0}{n}( (m_1-h_1,m_2-h_2,Z),(x,y, z)).
\end{equation}

Applying the same argument for the 2D case shows that the factorization of $K$ in \cref{eq:lowrank}
can be extended to the 3D case. More precisely, the Green function $G_0$ can be directly obtained as
\begin{equation}\label{eq:G03d}
  G_0(p,q) = \sum_{\ell\in\bbZ^3}\left(\Gamma(p-q+(\ell_1,\ell_2,2\ell_3Z)) 
  - \Gamma(p-q^\ast+(\ell_1,\ell_2,2\ell_3Z))\right),
\end{equation}
where $q^\ast=(q_1,q_2, 2Z+q_3)$ and $\Gamma$ is the Green function of the operator $-\Delta$ on the
whole space $\bbR^3$.  By taking $h=(h_1,h_2)$ and $z$ as parameters and using the similar notations
as the 2D case, one can reformulate \cref{eq:def_kappa3d} as
\begin{equation}
  \mu_{h}(m) \approx \int_{-Z}^{Z}(k_{h,z}*\eta_z)(m)\dd z 
  = \int_{-(Z-\delta)}^{Z-\delta}(k_{h,z}*\eta_z)(m)\dd z 
\end{equation}
where $*$ stands for the two-dimensional convolution with respect to the variables $m_1$ and $m_2$.
The same argument used in 2D case shows that one can factorize $k_{h,z}$ as
\begin{equation}
  k_{h,z}(m) \approx \sum_{\hat{h}} \sum_{\hat{z}} R_{h,\hat{h}} k_{\hat{h},\hat{z}}(m) R_{z,\hat{z}},
\end{equation}
by choosing proper interpolating sets $\{\hat{z}\}$ and $\{\hat{h}\}$, where $R_{h,\hat{h}}$ and
$R_{z,\hat{z}}$ are the interpolation operators in the $h$ and $z$ variables, respectively.

With the help of this approximation, \cref{eq:def_kappa3d} can be simplified as
\begin{equation}\label{eq:lowrank_app3d}
  \mu_h(m_1,m_2) \approx 
  \sum_{\hat{h}} R_{h,\hat{h}} \left( \sum_{\hat{z}} k_{\hat{h},\hat{z}}*
  \left(\int_{-(Z-\delta)}^{Z-\delta}  R_{z,\hat{z}}\eta_z\dd z\right)\right)(m_1,m_2).
\end{equation}
This results the same three-step procedure for effectively approximating the forward map, with the
minor differences that the interpolation over the $h$ variable is now two-dimensional and the
convolution is for 2D.

\begin{algorithm}[htb]
  \begin{small}
    \begin{center}
      \begin{algorithmic}[1]
        \Require $c, n_{\cnn}, n_{\cnn2} \in\bbN$, 
        $\mu\in\bbR^{N_{m_1}\times N_{m_2}\times N_{h_1}\times N_{h_2}}$
        \Ensure $\eta\in\bbR^{N_x\times N_y\times N_{{z}}}$
        \State Reshape $\mu$ to a three-dimensional tensor by vectorizing the last two-dimension
        and still denote it by $\mu$
        \State $\tilde{\mu} \leftarrow \encoding\trd[c](\mu)$
        \State $\tilde{\eta} \leftarrow \BCR\snd[c, n_{\cnn}](\tilde{\mu})$
        \State $\bar{\eta}\leftarrow \decoding\trd[N_z](\tilde{\eta})$
        \add{\State $\eta \leftarrow \CNNthree[w, n_{\cnn2}](\bar{\eta})$}
        \State \sffont{return} $\eta$
      \end{algorithmic}
    \end{center}
  \end{small} 
  \caption{\label{alg:eitinverse3d} Neural network architecture for the one-sided detection inverse map 
    $\mu\to \eta$ for 3D case.}
\end{algorithm}

Following the same reasoning for the 2D case, the NN architecture for the inverse map for the
one-sided detection in \cref{alg:eitinverse3d}. Below we briefly comment on the layers used,
focusing on the differences with the 2D case.
\begin{itemize}
\item Encoding module. $\tilde{\mu} = \encoding\trd[c](\mu)$ compresses the data
  $\mu\in\bbR^{N_{1}\times N_{2}\times N_3}$ to $\tilde{\mu}\in\bbR^{N_1\times N_2\times c}$ locally
  with respect to the first and second dimensions. 
  Similar as the 2D case, this layer can be implemented by a two-dimensional convolution $\Convtwo$
  with window size $1\times 1$ and channel number $c$ by taking the third dimension of $\mu$ as the
  channel direction. Noticing the compressed layer is essentially a restriction operator, we only
  use one \Convtwo layer with linear activation function.

\item $\BCR\snd$. In \cref{sec:forward}, the network $\BCR\first[c,n_{\cnn}]$ can be treated as a
  compact form of the \revised{fully connected}{full-width} convolutional layers with number of
  layer $n_{\cnn}$ and channel number $c$.  Here the $\BCR\snd[c,n_{\cnn}]$ is the compact form of
  the \revised{fully connected}{full-width} convolutional layers with number of layer $n_{\cnn}$
  and channel number $c$.  We refer readers to \cite{bcr} for more details.

\item Decoding module. $\eta=\decoding\trd[N_z](\tilde{\eta})$ extends the set of two-dimensional
  data $\tilde{\eta}\in\bbR^{N_{1}\times N_{2}\times c}$ to the three-dimensional data
  $\eta\in\bbR^{N_{1}\times N_{2}\times N_3}$. This module \delete{is the adjoint of the encoding
  module and} can be implemented by the two-dimensional convolutional layer \Convtwo with window
  size $1\times 1$ and channel number $N_z$ by one layer with linear activation function.
  
  \add{
\item $\CNNthree$. In \cref{subsec:inv1s}, $\eta=\CNNtwo[w,n_{\cnn}](\bar{\eta})$ is a
  post-processing module on the output of the decoding. $\eta=\CNNthree[w,n_{\cnn}](\bar{\eta})$
  which maps the data $\bar{\eta}\in\bbR^{N_x\times N_y\times N_z}$ to $\eta\in\bbR^{N_x\times
  N_y\times N_z}$, is the three-dimensional analog of $\CNNtwo$.  It is a three-dimensional
  convolutional neural network with $n_{\cnn}$ convolutional layers and $w$ as the window size. 
  ReLU is used as the activation function for all intermediate layers and no activation function is
  applied after the last layer.
  }
\end{itemize}

\paragraph{Two-sided detection.}
Following the 2D case, the two-sided detection can be treated as a combination of four one-sided
detections and a post-process to combine the four parts. The NN architecture is summarized in
\cref{alg:eitTS3d}.  Below we briefly comment on the two new layers used.
\begin{itemize}
\item {Concatenate layer.} $\eta\leftarrow \concatenate\snd(\eta_1,\eta_2,\eta_3,\eta_4)$
  concatenates the 3-tensors $\eta_i\in\bbR^{N_1\times N_2\times c}$, $i=1,2,3,4$ to a 3-tensor with
  size $\eta\in\bbR^{N_1\times N_2\times 4c}$ on the third direction.
\item {Convolutional layer.} $\eta\leftarrow \Convtwo[c, w](\eta)$ is the two-dimensional
  convolutional layer with channel number $c$ and window size $w$.
\end{itemize}

\begin{algorithm}[htb]
  \begin{small}
    \begin{center}
      \begin{algorithmic}[1]
        \Require $c, n_{\cnn}, n_{\cnn2}, n_{\cnn3}, w, w_2 \in\bbN$, 
        $\mu^{\pm\pm}\in\bbR^{N_{m_1}\times N_{m_2}\times N_{h_1}\times N_{h_2}}$ 
        \Ensure $\eta\in\bbR^{N_x\times N_y\times N_{{z}}}$
        \State Denote $\mu_{i}$, $i=1,2,3,4$ by all the cases of $\mu^{\pm\pm}$ 
        \For { $i$ from $1$ to $4$} 
        \State Reshape $\mu_i$ to a three-dimensional tensor by vectorizing the last
        two-dimension and still denote it by $\mu_i$ 
        \State $\tilde{\mu}_i \leftarrow \encoding\trd[c](\mu_i)$ 
        \State $\tilde{\eta}_i \leftarrow \BCR\snd[c, n_{\cnn}](\tilde{\mu}_i)$ 
        \EndFor 
        \State $\tilde{\eta}\leftarrow \concatenate\snd(\tilde{\eta}_1,\tilde{\eta}_2,\tilde{\eta}_3,\tilde{\eta}_4)$ 
        \For {$k$ from $1$ to $n_{\cnn3}$} 
        \State $\tilde{\eta}\leftarrow \Convtwo[4c, w_2](\tilde{\eta})$
        \EndFor 
        \State $\bar{\eta} \leftarrow \decoding\trd[N_z](\tilde{\eta})$ 
        \add{\State $\eta \leftarrow \CNNthree[w, n_{\cnn2}](\bar{\eta})$}
        \State \sffont{return} $\eta$
      \end{algorithmic}
    \end{center}
  \end{small} 
  \caption{\label{alg:eitTS3d} Neural network architecture for the two-sided detection inverse map
    $\mu^{\pm\pm}\to \eta$ for 3D case.}
\end{algorithm}

\subsection{Numerical results}\label{sec:numerical3d}

The setup of the neural network is almost the same as that for the 2D case in
\cref{sec:numerical2d}. The only difference is that the window size of the convolutional layers in
\cref{alg:eitinverse3d,alg:eitTS3d} is set as \revised{$3\times 3$}{$w=w_2=(3,3)$ rather than
  $w=w_2=3$ for the 2D case}.  For each sample of the training and test data sets, $\eta(x)$ is of
the form
\begin{equation}\label{eq:data3d}
  \eta(x) = \sum_{i=1}^{4} \rho 
  \exp\left( -\frac{1}{2}(x-c^{(i)})^\T(\Theta^{(i)})^{-1}(x-c^{(i)}) \right),
\end{equation}
where $\rho=1000$. The matrix $\Theta^{(i)}\in\bbR^{3\times 3}$ is generated with the eigenvalues
uniformly distributed in $[0.0125, 0.05]$ and the eigenvectors uniformly sampled from the unit
sphere $\bbS^2$. Similar as the 2D case, $Z = 1 / 4$. Two types of data are generated to test the
performance of the proposed NNs:
\begin{itemize}
\item Shallow inclusions. The location of Gaussians is
  $c^{(i)}\in\cU([0,1]\times[0,1]\times[0.05,0.2])$, which is used for the test of the one-sided
  detection.
\item Deep inclusions. The location of Gaussians is
  $c^{(i)}\in\cU([0,1]\times[0,1]\times[-0.2,0.2])$, used for the test of the two-sided detection.
\end{itemize}
The domain $\Omega$ is discretized by a $40\times 40\times 20$ Cartesian grid. Both the training
data and test data are generated by numerically solving \cref{eq:schrodinger}.

\begin{figure}[h!]
  \centering
  \includegraphics[width=0.4\textwidth,clip]{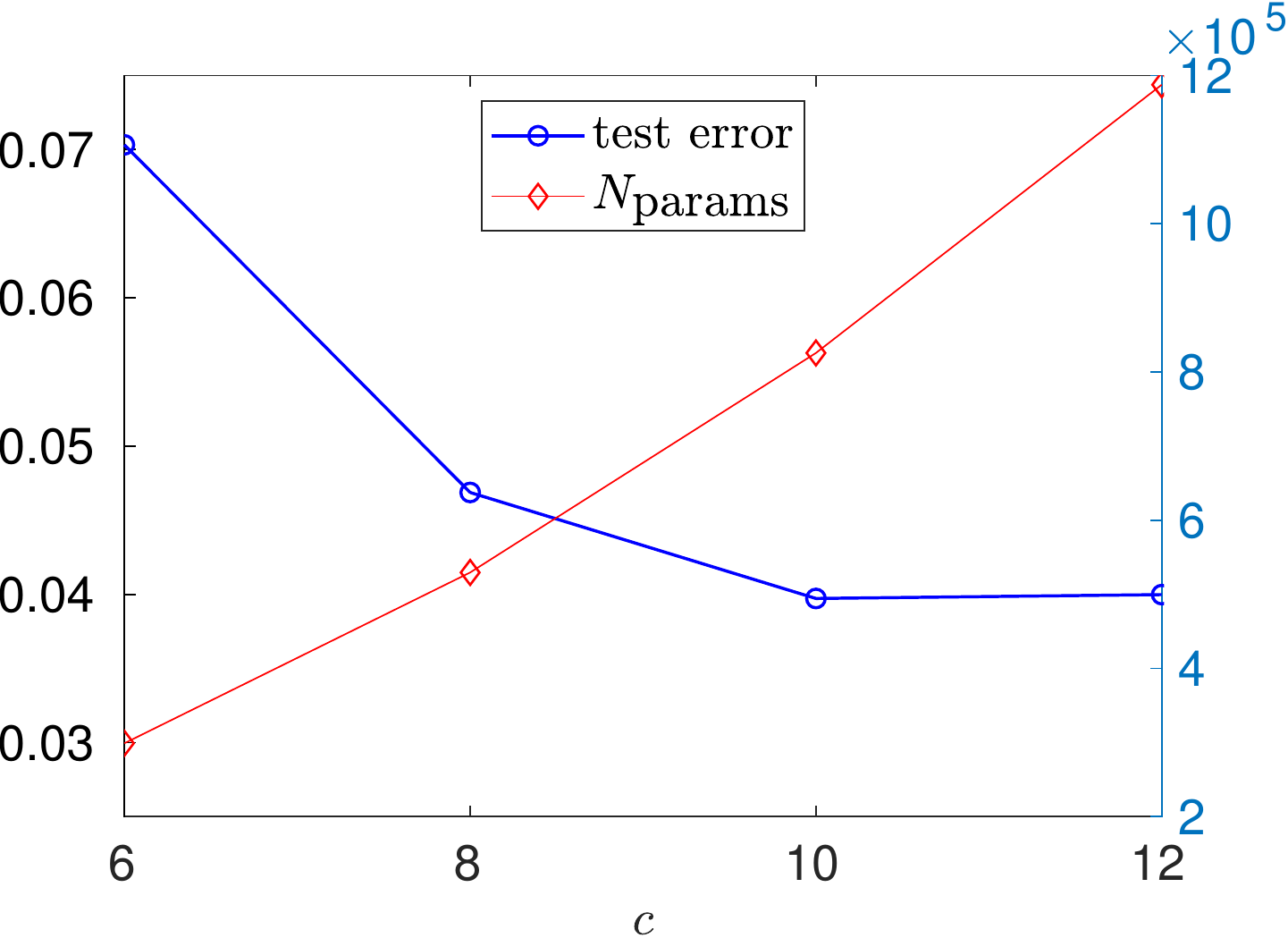}
  \caption{\label{fig:err3d}The relative error and the number of parameters for the one-sided
    detection inverse problem for different number of channels $c$ for 3D case.}
\end{figure}

\begin{figure}[h!]
    \centering
    \subfloat[$\mu$]{
    \includegraphics[width=0.3\textwidth]{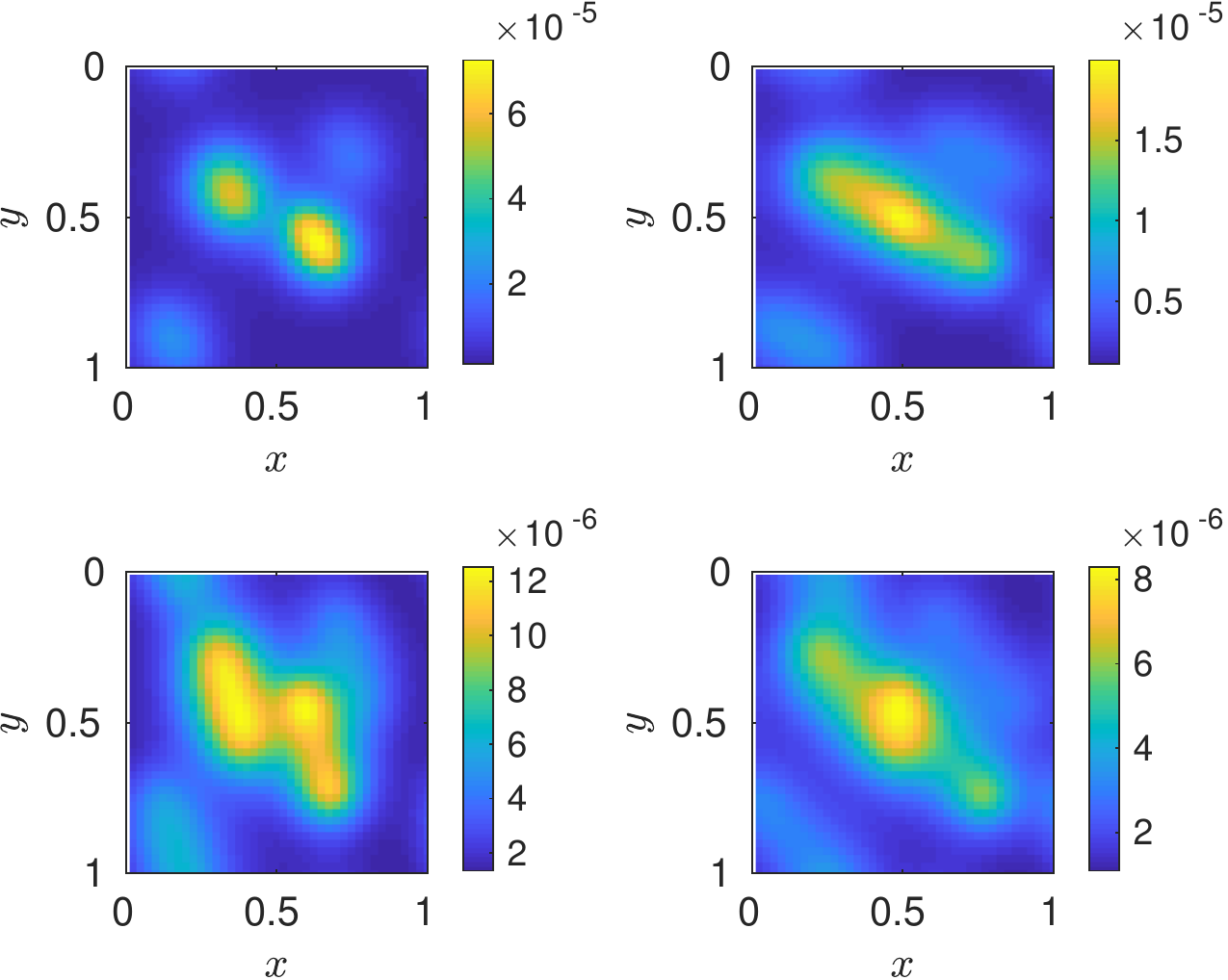}
    }\hspace{0.02\textwidth}
    \subfloat[$\eta$]{
    \includegraphics[width=0.3\textwidth]{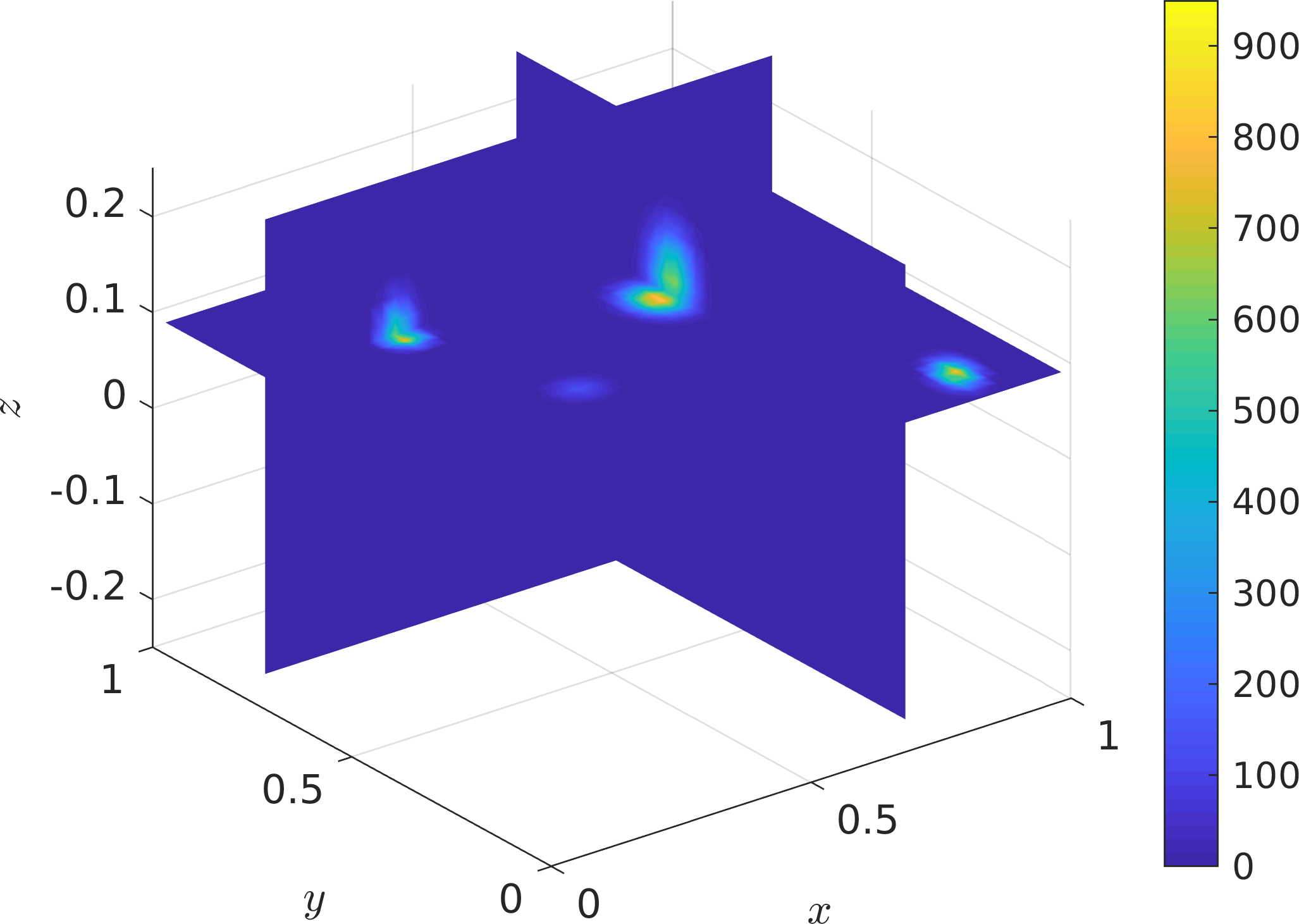}
    }\hspace{0.02\textwidth}
    \subfloat[$\eta_{NN}$]{
    \includegraphics[width=0.3\textwidth]{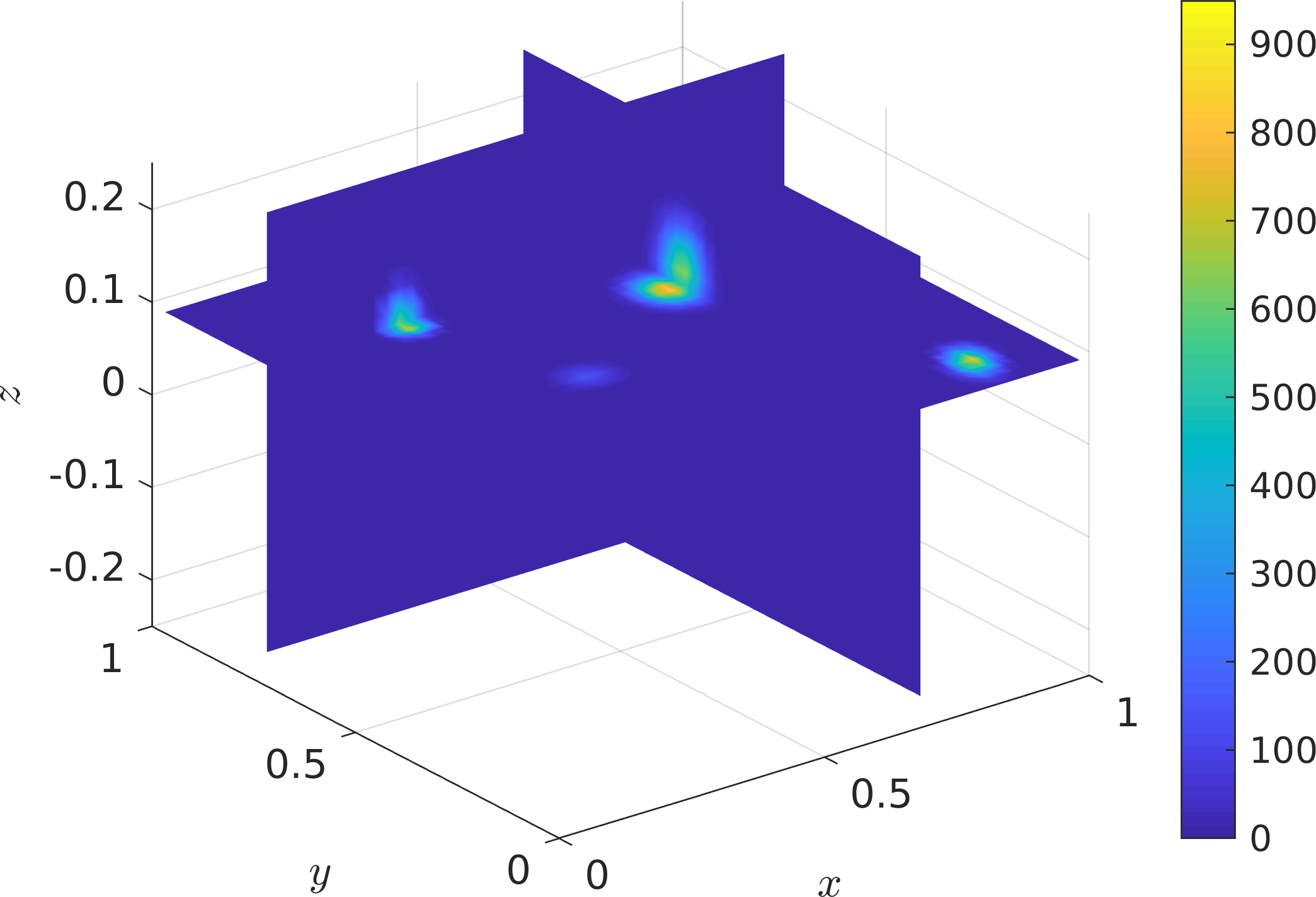}
    }
    \caption{\label{fig:fig3d} A sample in the test data for the one-sided detection inverse
    problem with $c=8$ on the shallow inclusions for 3D case. (a) Some slice of the
    data $\mu$;
    (b) The reference solution $\eta$; (c) The prediction $\eta_{NN}$. The relative error 
  of this sample is \revised{$5.6e-2$}{$5.2e-2$}.}
\end{figure}

\paragraph{One-sided detection for shallow inclusions.}
The NN for the inverse map in \cref{alg:eitinverse3d} is tested with shallow inclusions.  $10$K
pairs of $(\eta,\mu)$ are used to train the NN parameters and another $5$K pairs are reserved as the
test data. The data set is smaller compared to the 2D case due to the memory limitation of the current
GPUs. \Cref{fig:err3d} plots the test error and the number of parameters for different choices of
$c$. The test error is comparable with that of the 2D case. As $c$ increases, the error decays first
and then stagnates around $c=10$. \cref{fig:fig3d} illustrates the NN prediction and the reference
solution of a specific sample from the test data set. The plots indicate that the NN produces
accurate results in terms of the location, the shape, and the magnitude of the inclusions.

\begin{figure}[h!]
  \centering
  \includegraphics[width=0.4\textwidth,clip]{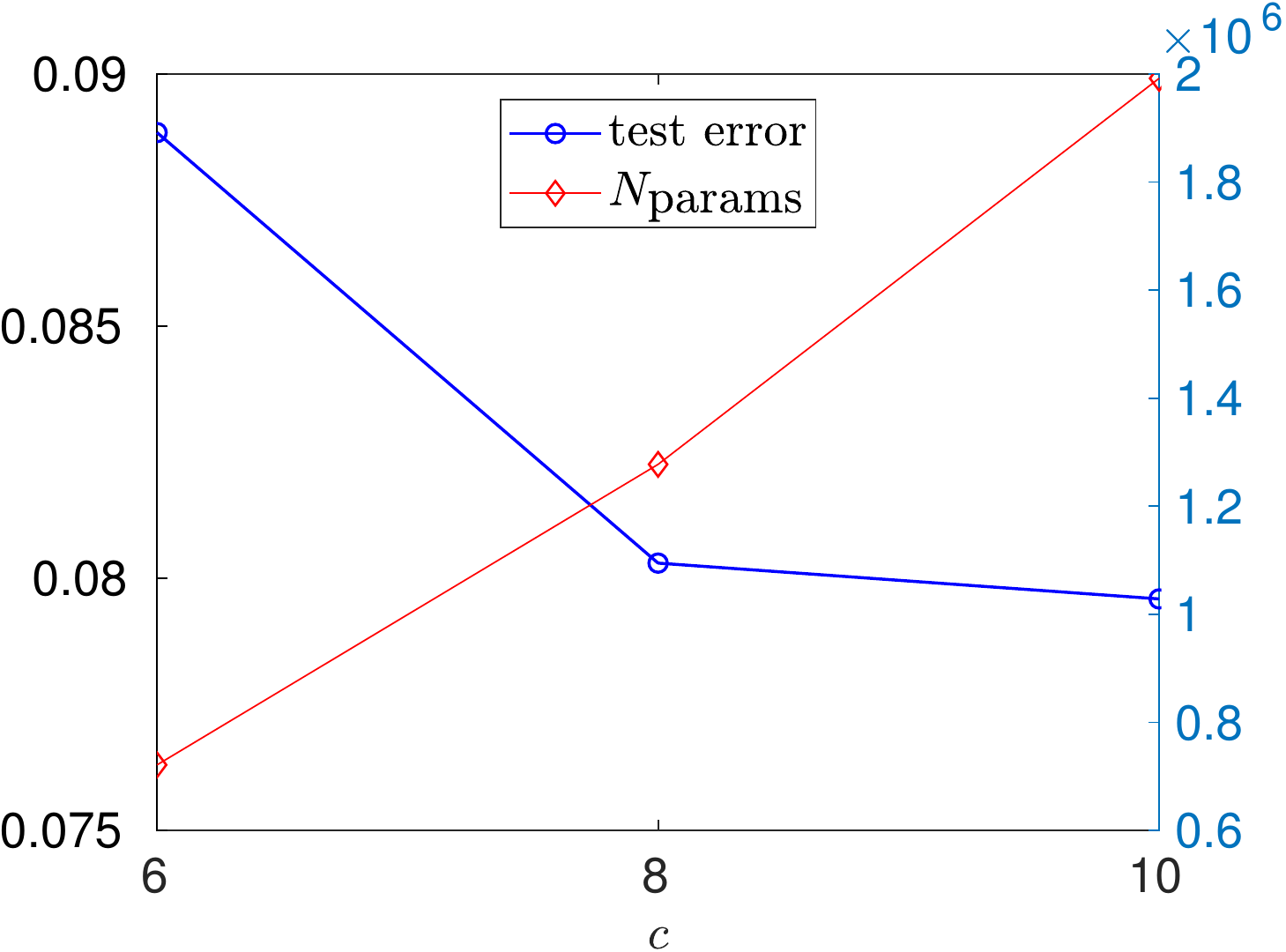}
  \caption{\label{fig:TSerr3d}The relative error and the number of parameters for the two-sided
    detection inverse problem for different number of channels $c$ for 3D case.}
\end{figure}

\begin{figure}[h!]
    \centering
    \subfloat[$\mu^{+,+}$]{
    \includegraphics[width=0.3\textwidth]{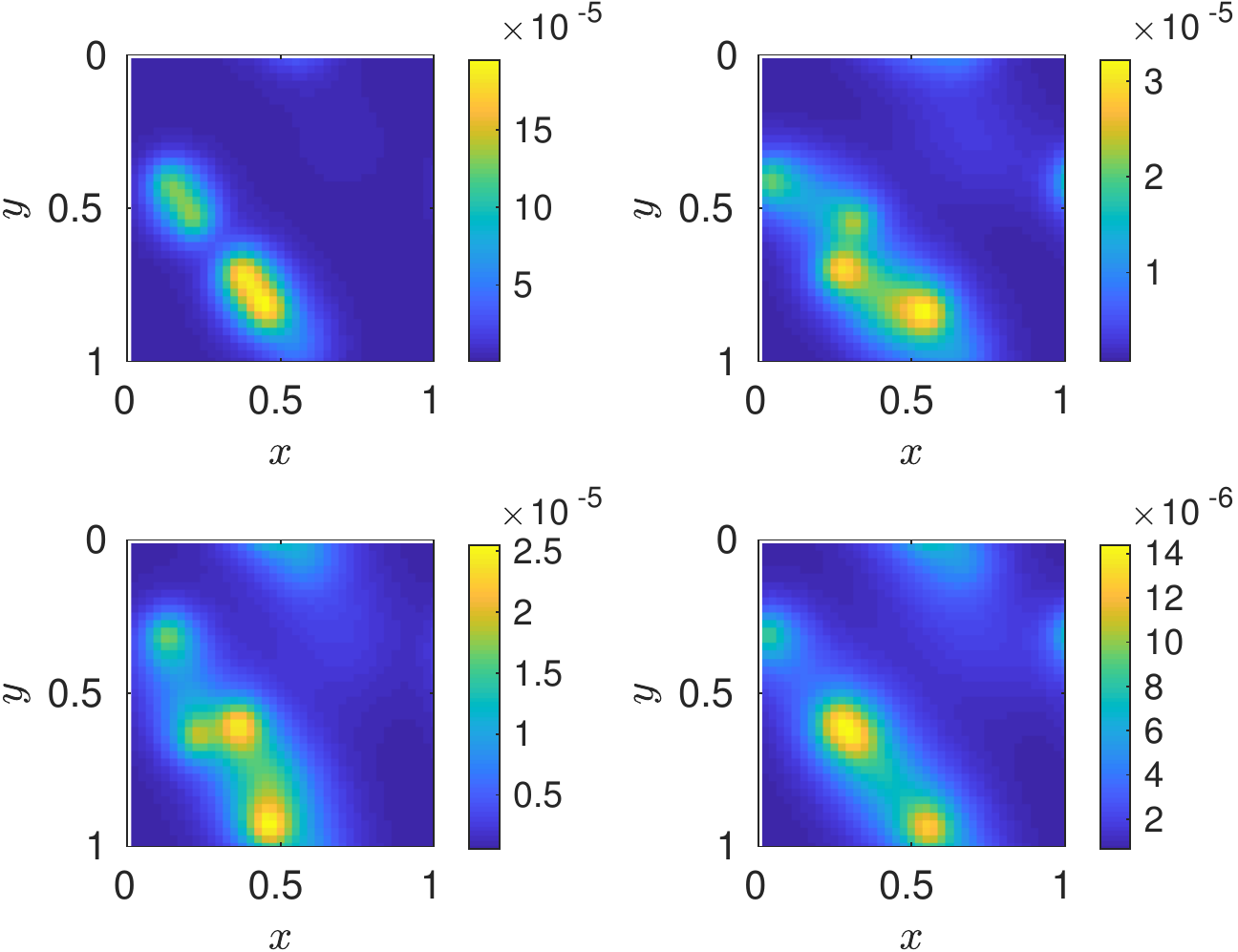}
    }\hspace{0.02\textwidth}
    \subfloat[$\eta$]{
    \includegraphics[width=0.3\textwidth]{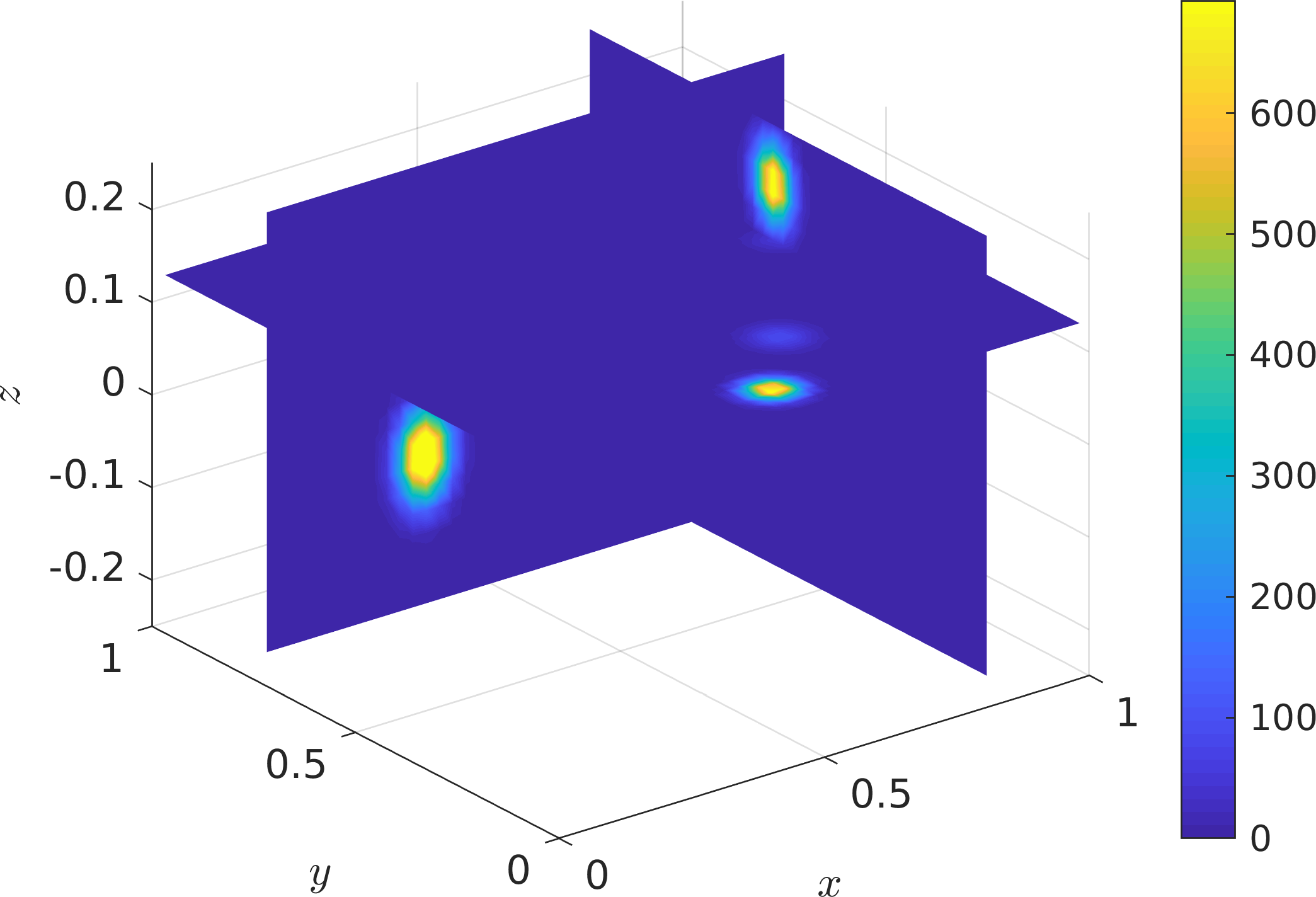}
    }\hspace{0.02\textwidth}
    \subfloat[$\eta_{NN}$]{
    \includegraphics[width=0.3\textwidth]{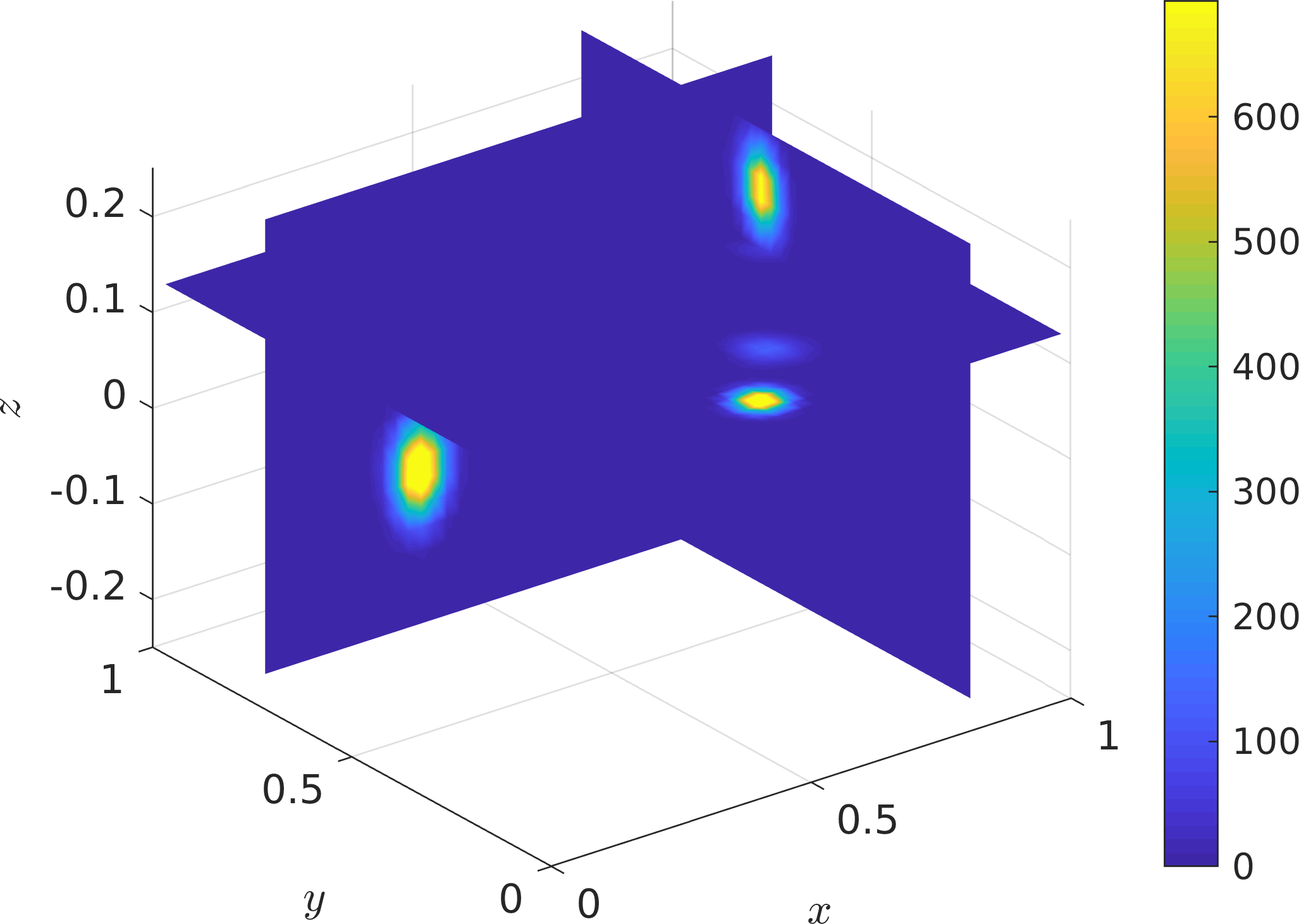}
    }
    \caption{\label{fig:TSfig3d} A sample in the test data for the two-sided detection inverse
    problem with $c=8$ on deep inclusions for 3D case. (a) Some slice of the
    data $\mu^{+,+}$;
    (b) The reference solution $\eta$; (c) The prediction $\eta_{NN}$. The relative error 
  of this sample is \revised{$9.0e-2$}{$8.1e-2$}.}
\end{figure}

\paragraph{Two-sided detection for deep inclusions.}
The neural network in \cref{alg:eitTS3d} for two-sided detection is tested with deep
inclusions. $8$K pairs of $(\eta,\mu)$ are used to train the neural network and another $2$K pairs
are reserved for testing. \cref{fig:TSerr3d} summarizes the test error and the number of parameters
for different values for $c$. The test accuracy is comparable with that of the 2D two-sided
detection case. \Cref{fig:TSfig3d} plots the prediction solution and the reference solution of a
specific sample in the test data.

\section{Conclusions}\label{sec:conclusion}

This paper proposes novel neural network architectures for EIT problems. Mathematically, these NNs
approximate the forward and inverse maps between the electrical conductivity and the resulting DtN
map. A perturbative analysis for the weak-inclusion regime suggests a dimension-reduction
approximation, which further inspires the NN architecture design.  Numerical results demonstrate
that the proposed NNs approximate the forward or inverse maps with reasonable accuracy.

Using neural networks for approximating the forward and inverse maps has several clear advantages.
First, once the neural networks are well trained, they can produce rather accurate prediction
efficiently; Second, the correct regularization for the inverse map can be automatically captured by
the neural network from the training set; Third, the neural networks proposed in this paper are
compact and easy to train, thus applicable to applications with rather limited data.

The discussion here focuses on the rectangle/cuboid domains.\delete{ with the periodic boundary
condition specified on the sides. It is not difficult to extend it to the rectangle domains with
other boundary conditions.} For the spherical domains, one can turn them into the rectangle/cuboid
configuration by resorting to the polar/spherical transformations. For arbitrary convex bounded
Lipschitz domains, it is possible to extend the current NN architectures by carefully
reparametrizing the domain, though it could require technical efforts.

\add{Recently in medical imaging applications, adversarial attack
  \cite{Szegedy2013IntriguingPO,antun2019instabilities,finlayson2019adversarial} has become an
  important issue for deep learning. As our NN architecture is quite compact, the resulting NN does
  not overfit the training data. It is expected that our NN could be less vulnerable to adversarial
  attacks. Detailed study in this direction would be part of future work. }

\section*{Acknowledgments}
The work of Y.F. and L.Y. is partially supported by the U.S. Department of Energy, Office of
Science, Office of Advanced Scientific Computing Research, Scientific Discovery through Advanced
Computing (SciDAC) program.  The work of L.Y. is also partially supported by the National Science
Foundation under award DMS-1818449.  This work is also supported by the GCP Research Credits Program
from Google and AWS Cloud Credits for Research program from Amazon.

\bibliographystyle{abbrv}
\bibliography{nn}
\end{document}